\documentclass[reprint,amsmath,amssymb,aps,superscriptaddress,prd,nofootinbib, floatfix]{revtex4-2}
\usepackage{geometry}
\usepackage[unicode=true,pdfusetitle,
bookmarks=true,bookmarksnumbered=true,bookmarksopen=true,bookmarksopenlevel=1,
breaklinks=false,pdfborder={0 0 0},backref=false,colorlinks=true]{hyperref}
\hypersetup{citecolor=blue,filecolor=blue,linkcolor=blue,urlcolor=blue,pdfauthor={Name}}

\usepackage{graphicx}
\usepackage{mathtools}
\usepackage{siunitx}
\usepackage{xcolor}
\usepackage[normalem]{ulem}
\usepackage{url}
\usepackage{soul}
\usepackage{booktabs}
\usepackage{multirow}

\newcommand{\vect}[1]{\ensuremath{\boldsymbol{#1}}}
\newcommand{\msun}{M_\odot}

\begin{document}

\title{A search technique to observe precessing compact binary mergers in the advanced detector era}

\author{Connor McIsaac}
\affiliation{University of Portsmouth, Portsmouth, PO1 3FX, United Kingdom}
\author{Charlie Hoy}
\affiliation{University of Portsmouth, Portsmouth, PO1 3FX, United Kingdom}
\author{Ian Harry}
\affiliation{University of Portsmouth, Portsmouth, PO1 3FX, United Kingdom}

\date{\today}

%----------------------------------------------------------
%----------------------- Abstract -------------------------
%----------------------------------------------------------
\begin{abstract}
  Gravitational-wave signals from compact binary coalescences are most efficiently identified through matched filter searches, which match the data against a pre-generated bank of gravitational-wave templates. Although different techniques for performing the matched filter, as well as generating the template bank, exist, currently all modelled gravitational-wave searches use templates that restrict the component spins to be aligned (or anti-aligned) with the orbital angular momentum. This means that current searches are less sensitive to gravitational-wave signals generated from binaries with generic spins (precessing), suggesting that, potentially, a significant fraction of signals may remain undetected. In this work we introduce a matched filter search that is sensitive to signals generated from precessing binaries and can realistically be used during a gravitational-wave observing run. We take advantage of the fact that a gravitational-wave signal from a precessing binary can be decomposed into a power series of five harmonics, to show that a generic-spin template bank, which is only $\sim 3\times$ larger than existing aligned-spin banks, is needed to increase our sensitive volume by $\sim 100\%$ for neutron star black hole binaries with total mass larger than $17.5\, M_{\odot}$ and in-plane spins $>0.67$. In fact, our generic spin search performs as well as existing aligned-spin searches for neutron star black hole signals with insignificant in-plane spins, but improves sensitivity by $\sim60\%$ on average across the full generic spin parameter space. We anticipate that this improved technique will identify significantly more gravitational-wave signals, and, ultimately, help shed light on the unknown spin distribution of binaries in the universe.
\end{abstract}
\keywords{gravitational waves, signal detection, overlapping signals}

\maketitle

\section{Introduction}

In the last 10 years, gravitational-wave observatories, such as Advanced LIGO, Advanced Virgo and KAGRA, have unlocked the gravitational-wave Universe~\cite{LIGOScientific:2014pky, VIRGO:2014yos, KAGRA:2020agh}.
At the time of writing, roughly 100 compact binary coalescences have been observed using data from these observatories by the LIGO-Virgo-KAGRA (LVK) collaborations~\cite{LIGOScientific:2018mvr, LIGOScientific:2020ibl, LIGOScientific:2021usb, LIGOScientific:2021djp}. In addition,
the public release of LVK data via the Gravitational Wave Open Science Center~\cite{LIGOScientific:2019lzm} has enabled external groups to analyse the data and identify additional events~\cite{Nitz:2018imz, Venumadhav:2019tad, Venumadhav:2019lyq, Nitz:2020oeq, Nitz:2021uxj, Olsen:2022pin}.

These many observations have been made possible by the development of complex search algorithms to matched filter the gravitational-wave data against a set of filter waveforms representing our best knowledge of the gravitational-wave signal emitted by compact binary mergers~\cite{Allen:2005fk, Cannon:2011vi, Babak:2012zx, Cannon:2012zt, DalCanton:2014hxh, Adams:2015ulm, Usman:2015kfa, Cannon:2015gha, Messick:2016aqy,Venumadhav:2019tad, Sachdev:2019vvd, Chu:2020pjv, Davies:2020tsx, Aubin:2020goo}.
However, while these searches have been undeniably successful, they are all limited in one important regard: They are all restricted to performing modelled searches with template banks that only contain aligned-spin templates, where the spin angular momentum of the two compact objects and the orbital angular momentum of the binary are aligned~\footnote{We note that in addition to matched-filter searches, unmodelled, or semi-modelled, pipelines have been developed to target the observation of sources that don't match well to our template waveforms~\cite{Klimenko:2008fu, Lynch:2015yin, Skliris:2020qax}. However, while these can match the sensitivity of matched-filtering when searching for high-mass signals, the performance of such searches does not match matched-filtering when considering systems with relatively low chirp mass, as we will do in this work.}.
This is due to the simple relationship between the sky position and orientation of aligned-spin binaries and the waveform observed by the gravitational wave detector, which allows us to analytically maximise over these extrinsic parameters~\cite{Allen:2005fk, Babak:2012zx}.

In this work, we revisit the problem of searching for compact binary coalescences where the spins are misaligned with the orbital angular momentum. When the spins are misaligned, spin-orbit coupling will cause the orbital angular momentum and the spin angular momenta to precess around the direction of the total angular momentum~\cite{Apostolatos:1994mx}. This effect will modulate the phase and amplitude of the observed gravitational waves, with the exact form of the modulation depending on the orientation and sky position of the binary. This dependence means that we can no longer analytically maximise over these extrinsic parameters, making the inclusion of these effects in a modelled search challenging.

There are two main formation channels for the production of compact binaries, through the isolated evolution of a pair of massive stars, or through the dynamical formation of binaries in dense stellar environments~\cite{Mandel:2018hfr, Mapelli:2018uds, Mapelli:2021taw, Gerosa:2021mno}.
For compact binary coalescences that evolved from a pair of massive binary stars, we expect the spin angular momenta of the components to be roughly aligned with the orbital angular momentum, with some misalignment present due to kicks caused during core collapse of either component~\cite{Kalogera:1999tq, Gerosa:2018wbw}.
For binaries that are formed through dynamic capture in dense stellar environment we expect to see isotropic spin distributions~\cite{Rodriguez:2016vmx}.
Observation of compact binary coalescences with large misaligned spins, or lack thereof, will therefore allow us to test the rate of binaries produced by different formation channels~\cite{Gompertz:2021xub, Stevenson:2017dlk} and is therefore of great astrophysical interest.
However, if current search methods are missing strongly precessing signals then this could introduce a bias to these measurements.
It is therefore important that we develop new methods in order to efficiently search for precessing signals.

When using only aligned-spin templates it has been demonstrated that reasonable sensitivity is retained to templates with moderate effects due to precession~\cite{Harry:2013tca,CalderonBustillo:2016rlt}, losing $\sim 17 - 23\%$ of our sensitivity in the neutron-star--black-hole parameter space compared to an ideal search~\cite{Harry:2013tca}. 
Aligned-spin templates are particularly effective when the component masses are close to equal, the magnitude of the spin angular momentum is small, or the orbit of the binary is close to face-on ($\iota=0$) or face-off ($\iota=\pi$).
In these cases the precession of the binary will only weakly effect the waveform, making it difficult to infer the presence or absence of precession in the observed signal~\cite{Green:2020ptm}.
For most individual events observed so far there are only weak constraints on the size of the misaligned spin components \cite{LIGOScientific:2020ibl, LIGOScientific:2021djp}.
Recently, strong evidence for precession has been claimed in one observed compact binary coalescence -- GW200129\_065458~\cite{Hannam:2021pit}. However, there may be some uncertainty in this measurement due to non-Gaussian noise at the time of the event and the uncertainty in the glitch model used to remove this noise~\cite{Payne:2022spz}.
In short, most of the compact binary mergers observed so far show no evidence for precession. Wile this is likely reflective of the underlying population, it is possible that precessing signals remain undetected in our data as we are only looking with aligned-spin waveforms.

In the past, several methods have been proposed to search for precessing signals~\cite{Apostolatos:1995pj, Apostolatos:1996rf, Buonanno:2002fy, Pan:2003qt, Buonanno:2004yd, Harry:2011qh, Harry:2016ijz}. The first set of these methods \cite{Apostolatos:1995pj, Apostolatos:1996rf, Buonanno:2002fy} considered the addition of phenomenological parameters in order to mimic the effects of precession on the observed waveform. 
This allowed for a lot of unphysical freedom in the models, increasing the observed noise. A method was proposed in \cite{Harry:2016ijz} explicitly adding all of the required parameters to describe the observed signals to the template bank. However, when we attempted to apply this technique and generate a set of filter templates covering a physically meaningful parameter space for neutron-star--black-hole signals we generated over $30{,}000{,}000$ templates without any sign of the generation process converging, making observation impractical.
However, such a method could be viable for other regions of the parameter space where less templates are required, such as high mass binaries.
In~\cite{Buonanno:2002fy} the authors suggested the use of a physical decomposition of the precessing waveform that could be used to maximise over extrinsic parameters of the binary, which was then developed in~\cite{Pan:2003qt, Buonanno:2004yd, Fazi:2009ifa, Harry:2011qh}. Although a method was developed to restrict the observed signals to physical regions of the parameter space, this step was computationally expensive~\cite{Fazi:2009ifa}, and an unconstrained statistic was used in practice, increasing the observed noise rate. An extension of this method to a targeted coherent search was proposed in~\cite{Harry:2011qh}, however, this also faced issues due to the increased noise background. Due to the difficulties inherent in each of these methods, no search incorporating waveforms with precessional effects in the search templates has been applied to Advanced LIGO, Advanced Virgo or KAGRA data.

In this work we introduce a new method that is similar in nature to~\cite{Buonanno:2002fy} but uses the harmonic decomposition proposed in \cite{Fairhurst:2019vut} to reduce the complexity of each of the templates. We show that we can minimise the unphysical freedom introduced by the maximisation over extrinsic parameters by using a subset of the available harmonics, while still recovering the majority of signal power from precessing events. We demonstrate that a precessing bank containing only 355160 templates not only increases the observed signal-to-noise ratio for precessing injections compared to an aligned-spin bank, but also increases the sensitive volume by $\sim 100\%$ for binaries with total mass larger than $17.5\, M_{\odot}$ and in-plane spins $>0.67$.

We will begin by reviewing the effect of precession on the evolution of compact binary coalescences and their signals in section \ref{sec:precession}. We review previous attempts to search for precessing compact binary mergers in section~\ref{sec:prec_search_methods}. We will then review the harmonic decomposition in section \ref{sec:harmonics} and its use in modelling the precessing signal for different sky positions and orientations. In section \ref{sec:search_motivation} we motivate why the harmonic decomposition offers a way to solve the precessing search problem. In section \ref{sec:precessing_search} we introduce our new modelled search using the harmonic decomposition to maximise over the set of intrinsic parameters. Finally, in section \ref{sec:harmonic_coinc} we demonstrate that with an appropriate choice of detection statistic we can improve the sensitivity of modelled searches to neutron-star--black-hole signals by $\sim60\%$ on average across the full generic spin parameter space. 

\section{\label{sec:precession}Gravitational-wave signals of precessing binaries}

In this section we provide an overview of the gravitational-wave signals produced by precessing binaries. For further details, we refer the reader to Refs.~\cite{Apostolatos:1994mx,Kidder:1995zr,Apostolatos:1995pj,Buonanno:2002fy}.

A binary consisting of two compact objects will slowly inspiral due to the emission of gravitational waves. The emitted gravitational waves carry away angular momentum from the binary along the direction of the orbital angular momentum $\vect{L}$. Assuming a quasi-circular orbit, the binary's evolution can be fully described by 8 parameters: the masses, $m_{1}$ and $m_{2}$, and the spin angular momentum vector, $\vect{S}_{1}$ and $\vect{S}_{2}$, of each compact object.

\begin{figure}
    \centering
    \includegraphics[width=0.45\textwidth]{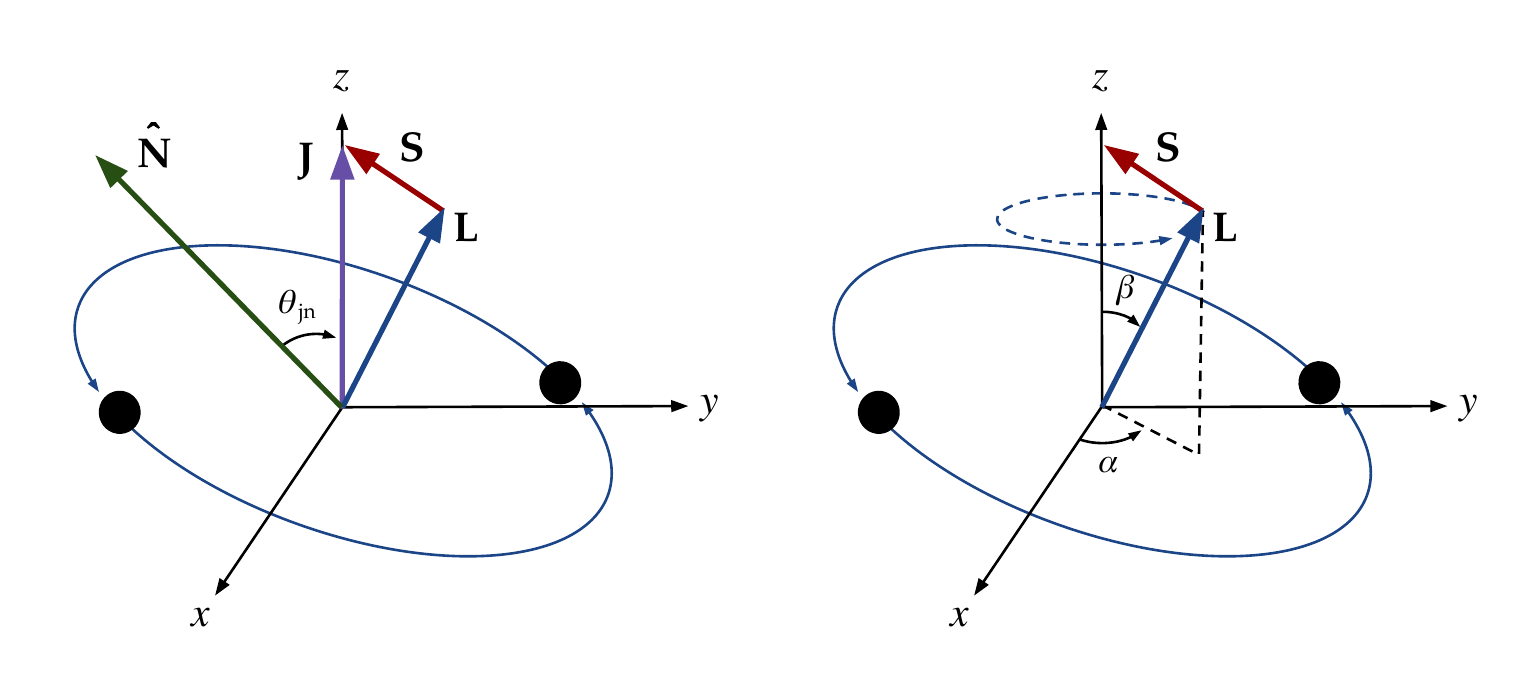}
    \caption{Illustration of a binary with spins misaligned with the orbital angular momentum demonstrating the angles $\alpha$, $\beta$ and $\theta_{\mathrm{JN}}$. In both panels the $z$ axis is aligned with the total angular momentum. The solid blue arrow represents the orbital angular moment, the solid red arrow represents the combined spin angular momentum, and the solid purple arrow represents the total angular momentum. The dashed blue arrow illustrates the path of the orbital angular momentum vector as the binary precesses. The solid green arrow shows the direction to the observer.}
    \label{fig:precession_jz}
\end{figure}

When $\vect{S}_1$ and/or $\vect{S}_2$ are non-zero and aligned or anti-aligned with $\vect{L}$, an `aligned-spin binary', spin-orbit and spin-spin couplings alter the rate of inspiral of the binary, adding a contribution to the overall phase of the observed signal, as well as the amplitude of the emitted gravitational waves~\cite{Kidder:1995zr,Campanelli:2006uy}. If the total spin angular momentum $\vect{S} = \vect{S}_{1} + \vect{S}_{2}$ is misaligned with $\vect{L}$ the binary will undergo spin-induced orbital precession~\cite{Apostolatos:1994mx}. For the case when $|\vect{L}| \ll |\vect{J}|$, $\vect{L}$, $\vect{S}_{1}$ and $\vect{S}_{2}$ precess around the approximately constant total angular momentum $\vect{J} = \vect{L} + \vect{S}$~\cite{Apostolatos:1994mx}, as illustrated in figure \ref{fig:precession_jz}. Although the emitted gravitational-waves continue to carry away angular momentum along $\vect{L}$, the precession of $\vect{L}$ around $\vect{J}$ implies that on average the angular momentum is emitted parallel to $\vect{J}$, with any emission orthogonal to $\vect{J}$ averaging to zero.

To simplify the modelling of a precessing binary, we utilize a source frame relative to the approximately constant vector $\vect{\Hat{J}}$. We define this frame such that the $z$ axis is parallel to the vector $\vect{\Hat{J}}$ and the $x$ axis is parallel to $\vect{\Hat{J}} \times \vect{\Hat{N}}$, where $\vect{\Hat{N}}$ is the direction to the observer. The angle between the vectors $\vect{\Hat{J}}$ and $\vect{\Hat{N}}$ will be given by
\begin{equation}
    \cos \theta_{\mathrm{JN}} = \vect{\Hat{J}} \cdot \vect{\Hat{N}}.
\end{equation}

Following works such as~\cite{Apostolatos:1994mx} we then define two angles to track the precession of $\vect{L}$, the phase of the precession, $\alpha$, and the opening angle $\beta$, both illustrated in figure \ref{fig:precession_jz}. The opening angle is given by
\begin{equation}
\label{eq:beta}
    \tan \beta = \frac{S_{\bot}}{|\vect{L}| + S_{\parallel}},
\end{equation}
where $S_{\parallel}$ and $S_{\bot}$ are the magnitudes of $\vect{S}$ parallel and perpendicular to $\vect{L}$ respectively.

As the precessing binary inspirals, $|\vect{L}|$ decreases while the magnitudes of the spin angular momenta $|\vect{S}_1|$ and $|\vect{S}_2|$ remain constant. Since $S_{\parallel}$ and $S_{\bot}$ also remain approximately constant throughout the inspiral, we see from equation~\ref{eq:beta} that the opening angle increases as the binary evolves. However, the rate of change of $\beta$ will be small compared to the precession frequency $\Omega_p = \Dot{\alpha}$~\cite{Brown:2012gs}.

In general the orbital frequency $\Omega_{\text{orb}}$ of the binary will be much larger than the precession frequency $\Omega_p$, such that the binary can complete several orbits before $\vect{L}$ changes significantly~\cite{Apostolatos:1994mx}. For this case we can approximate the dynamics of the binary as a set of quasi-circular orbits within an orbital plane which is precessing. This condition is called the ``adiabatic limit''.

We can gain some insight into the effect of precession on the observed signal by defining an instantaneous orbital plane that is perpendicular to $\vect{L}$ and modelling the dynamics within this plane using an aligned-spin waveform~\cite{Buonanno:2002fy}. We can then examine how the observed signal changes with a change in $\vect{L}$.

\begin{figure}
    \centering
    \includegraphics[width=0.45\textwidth]{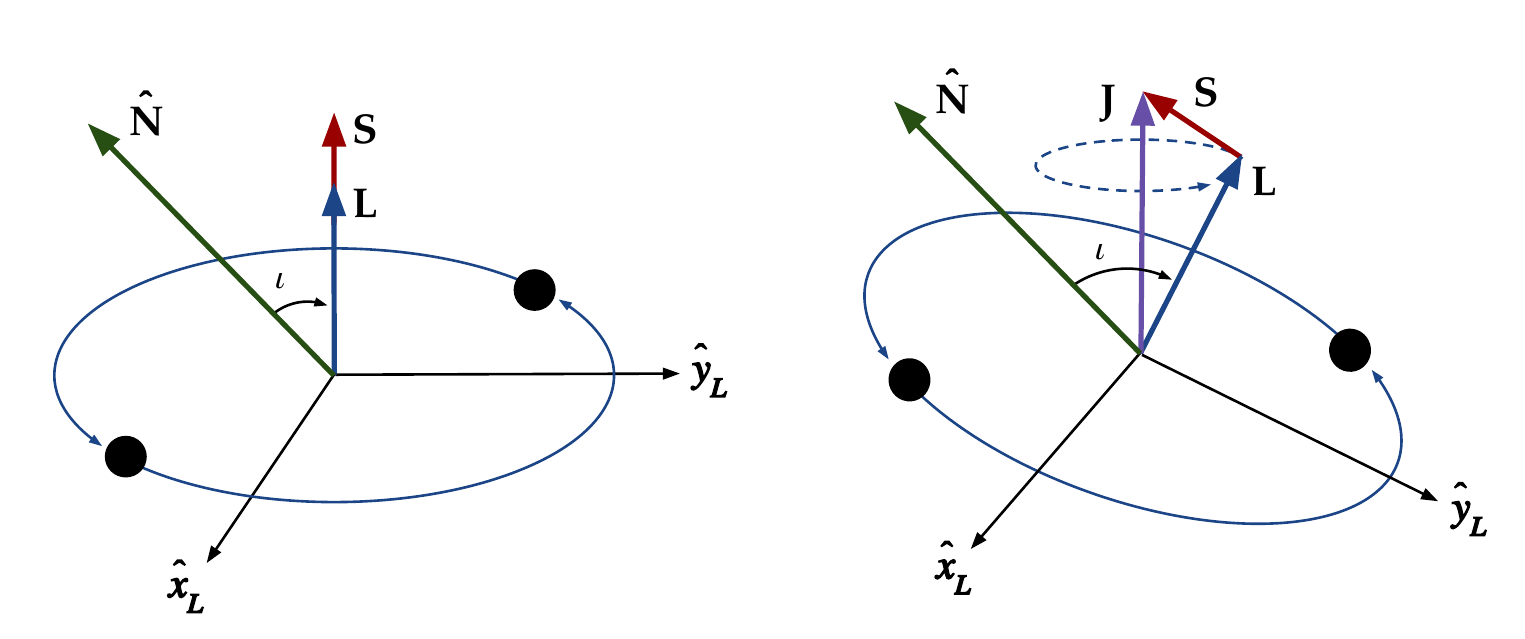}
    \caption{Illustration of a binary with spins aligned with the orbital angular momentum (left) and misaligned with the orbital angular momentum (right). The vectors $\vect{\Hat{x}}_L$ and $\vect{\Hat{y}}_L$ are defined to be orthogonal to $\vect{L}$ and form a basis for the instantaneous orbital plane of the binary. The solid blue arrow represents the orbital angular moment, the solid red arrow represents the combined spin angular momentum, and the solid purple arrow represents the total angular momentum. The dashed blue arrow illustrates the path of the orbital angular momentum vector as the binary precesses. The solid green arrow shows the direction to the observer.}
    \label{fig:precession_lz}
\end{figure}

We can define the vectors $\vect{\Hat{x}}_L$ and $\vect{\Hat{y}}_L$ to form a basis for the instantaneous orbital plane. Although $\vect{\Hat{x}}_L$ and $\vect{\Hat{y}}_L$ must be perpendicular to $\vect{L}$ we have the freedom to rotate them around $\vect{L}$, which will be degenerate with a change of the orbital phase. We choose $\vect{\Hat{x}}_L$ to be perpendicular to $\vect{\Hat{J}}$, giving us
\begin{equation}
    \vect{\Hat{x}}_L = \frac{\vect{\Hat{L}} \times \vect{\Hat{J}}}{\sin \beta}
    , \quad \quad
    \vect{\Hat{y}}_L = \frac{\cos \beta \vect{\Hat{L}} - \vect{\Hat{J}}}{\sin \beta},
\end{equation}
in the case that $\vect{\Hat{L}}$ and $\vect{\Hat{J}}$ are aligned we will choose $\vect{\Hat{x}}_L$ and $\vect{\Hat{y}}_L$ to be aligned with the $x$ and $y$ axes of the source frame. These vectors are illustrated in figure \ref{fig:precession_lz}.

For an aligned-spin binary inclined to the observer with angle $\iota$, where $\iota$ is defined by
\begin{equation}
    \label{eq:cosi}
    \cos \iota  = \vect{\Hat{L}} \cdot \vect{\Hat{N}},
\end{equation}
the two polarisations of the emitted gravitational waves due to the dominant quadrupole can be written as
\begin{gather}
    \label{eq:aligned_hp}
    h_{+}(t) = \frac{1 + \cos^2 \iota}{2r} A(t) \cos \left( 2 \phi(t) + 2 \phi_0 \right) \\
    \label{eq:aligned_hc}
    h_{\times}(t) = \frac{\cos \iota}{r} A(t) \sin \left( 2 \phi(t) + 2 \phi_0 \right).
\end{gather}
Here the amplitude, $A(t)$, and orbital phase, $\phi(t)$, can be calculated using the post-Newtonian formalism \cite{Blanchet:2006zz}. The orbital phase is defined as the angle between the orbital separation vector, $\vect{\Hat{r}}$, which points from $m_1$ to $m_2$, and the $x$ axis
\begin{equation}
    \label{eq:orbital_phase_def}
    \vect{\Hat{r}} = \vect{\Hat{x}_L} \cos \phi(t) + \vect{\Hat{y}_L} \sin \phi(t).
\end{equation}
The gravitational-wave detector observes a gravitational-wave, $h(t)$, which is a combination of the two polarisations, 
\begin{equation}
    \label{eq:aligned_response}
    h(t) = F_{+}(\Theta, \Phi, \Psi) h_{+}(t) + F_{\times}(\Theta, \Phi, \Psi) h_{\times}(t),
\end{equation}
where $F_{+}(\Theta, \Phi, \Psi)$ and $F_{\times}(\Theta, \Phi, \Psi)$ defines the detector's response to each of the two polarizations, and they depend on the orientation of the detector, as defined by the angles $\Theta, \Phi$ and $\Psi$. We can therefore write the observed gravitational-wave signal as
\begin{equation}
    \label{eq:aligned_h}
    h(t) = \frac{1}{D_{\mathrm{eff}}} A(t) \cos \left( 2\phi(t) + 2\phi_0 \right),
\end{equation}
where
\begin{equation}
    \label{eq:aligned_deff}
    D_{\mathrm{eff}} = r
    \left[ F_{+}^2
    \left( \frac{1 + \cos^2 \iota}{2} \right)^2
    + F_{\times}^2 \cos^2
    \iota \right]^{-1/2}
\end{equation}
and
\begin{equation}
    \label{eq:aligned_phase}
    2\phi_0 = 2\phi_c - \tan^{-1} \left( 2 \frac{F_{\times}}{F_{+}} \frac{\cos \iota}{1 + \cos^2 \iota} \right).
\end{equation}

As $\vect{L}$ precesses around $\vect{J}$ the orientation of the instantaneous orbital plane to the observer will change and $\iota$ will become time-dependent. The overall amplitudes of $h_{+}$ and $h_{\times}$ are both dependent on the inclination and are maximised when the binary is face-on ($\iota=0$) and minimised when edge-on ($\iota=\pi/2$). A change in the inclination also changes the relative amplitudes of the two polarisations; the observed gravitational waves will be circularly polarised when face-on and linearly polarised when edge-on. This time-dependent change in the inclination will therefore produce a modulation effect on both the amplitude and phase of the observed signal.

The rate of change of the inclination angle will be dependent on the viewing angle of the observer. For example, if the observer's line-of-sight is parallel to $\vect{J}$ then as $\vect{L}$ precesses around $\vect{J}$, the inclination angle will remain constant, while for an observer whose line-of-sight is initially parallel to $\vect{L}$ the inclination will oscillate between $\iota = 0$ and $\iota = 2 \beta$ with a frequency of $\Omega_p$. If $\beta$ is large this will produce a strong modulation effect.

In the aligned-spin case, we can see in equations \ref{eq:aligned_hp} and \ref{eq:aligned_hc} that the gravitational wave phase is given by twice the orbital phase $\phi(t)$, which is simply the accumulated phase due to the orbital frequency $\Omega_{\text{orb}}$. However, when the orbital plane is precessing, $\vect{\Hat{x}_L}$ will also rotate around $\vect{\Hat{L}}$ and the evolution of $\phi(t)$ becomes dependent on both the orbital and precession frequencies. This relationship is given by~\cite{Arun:2008kb}
\begin{equation}
    \label{eq:precessing_phase}
    \Dot{\phi} = \Omega_{\text{orb}} - \Omega_p \cos \beta
\end{equation}
introducing another modification to the phase of the observed signal.

\section{Previous methods to search for precessing signals}
\label{sec:prec_search_methods}
In the case of an aligned-spin system we can parameterise the binary with two component masses, $m_1$ and $m_2$, two spin spin magnitudes, $s_{1z}$ and $s_{2z}$, two angles describing the orientation of the binary, $(\iota, \phi_0)$, and three angles describing the orientation of the detector $(\Theta, \Phi, \Psi)$. Only the two masses and spins contribute to the evolution of the observed signal, while the five angles produce constant amplitude and phase factors given by equations \ref{eq:aligned_deff} and \ref{eq:aligned_phase}. This allows us to analytically maximise over the five angles using a phase-maximised matched filter~\cite{Allen:2005fk}, leaving only the two masses and spins to maximise over. This is done using a large set of filter waveforms chosen so as to sufficiently cover the full range of masses and spins, which we refer to as a ``template bank''.

In the case of a precessing binary we once again have two component masses, $m_1$ and $m_2$, however, we now have six spin components, $\vect{S}_1$ and $\vect{S}_2$. We use three angles to define the orientation of the binary, two to define the initial orientation of the orbital plane, $(\theta_{\mathrm{JN}}, \alpha_0)$, where $\alpha_0$ is the initial precession phase, and one to define the initial orbital phase, $\phi_0$~\footnote{We note that $\alpha_0$ is specified completely by $\vect{S}_1$ and $\vect{S}_2$ and the masses and is not an additional degree of freedom}. Finally we require the three angles describing the orientation of the detector $(\Theta, \Phi, \Psi)$. In this case the two polarisations, $h_+$ and $h_{\times}$ have a more complex dependence on $\theta_{\mathrm{JN}}$ than just a phase an amplitude shift and the polarizations themselves are no longer related by a simple phase shift. Therefore, in addition to the four extra spin components, one also needs to consider the effect of $\theta_{\mathrm{JN}}$, $\Psi$ and $\phi_0$ when developing a search to target precessing binaries.

Several methods have been proposed to tackle the problem of searching for precessing compact binary mergers. In~\cite{Apostolatos:1995pj, Apostolatos:1996rf} a small set of new parameters were introduced that modulate the phase of a non-precessing waveform in order to mimic the effects caused by precession. However, adding these parameters to the template bank is computationally expensive and the resulting templates do not provide adequate match with precessing waveforms~\cite{Grandclement:2002dv}. This was then extended in~\cite{Buonanno:2002fy} where the authors added a set of parameters that modulate the amplitude and phase of the waveform. The precessing signal is then expressed as a combination of modulated waveforms each with a different amplitude and phase. The signal-to-noise ratio is then maximized analytically for the amplitude and phase of each modulated waveform, leaving only a few parameters to be added to the template bank. However, the maximisation over the amplitudes and phases of the modulated waveforms allows for many unphysical combinations and the improvement in the match with precessing signals is outweighed by an increase in the observed noise due to the increased parameter space~\cite{VanDenBroeck:2009gd}.

Another method was proposed in~\cite{Buonanno:2002fy} and developed in~\cite{Pan:2003qt, Buonanno:2004yd}, in which the binary and its precession are modelled using a single spin $\vect{S}_1$ using only the $l=2$, $|m| = 2$ harmonics in the instantaneous orbital plane. The gravitational wave polarisations, $h_{+}$ and $h_{\times}$, in the radiation frame are re-expressed as a combination of five waveforms, $Q^I$, using the $l=2$ spherical harmonics as a basis. These five waveforms only depend on the intrinsic parameters of the binary and a constant time and phase offset. The observed signal is then expressed as a combination of the five waveforms, each multiplied by a coefficient, $P_I$, which is dependent on the sky position and orientation of the binary.

Maximising over the sky location and orientation of the binary is then achieved by maximising over the coefficients $P_I$. Although the five components of $P_I$ depend on six parameters in the case of a single detector, the four parameters describing the location and orientation of the detector $(\Theta, \Phi, \Psi, r)$ only enter as two independent combinations. The five components of $P_I$ therefore depend on only four independent parameters and the values of $P_I$ must have some constraint. In~\cite{Pan:2003qt} a method was developed to maximise the signal-to-noise ratio over the constrained values of $P_I$, however this is computationally expensive and an unconstrained signal-to-noise ratio maximisation of $P_I$ is used instead, allowing the components to take unphysical values and increasing the rate of noise triggers. The constraint problem becomes more difficult when we consider multiple detectors. For the 2-detector example we have the five $P_I$ components measured independently in two detectors, but these 10 $P_I$ values still only depend on 6 physical parameters, and so if these values are not constrained considerable unphysical freedom is allowed.
This method was extended to a targeted coherent search in~\cite{Harry:2011qh}. In this work the authors identified areas of the parameter space where precession effects were weak and restricted waveforms in these areas to only the dominant $Q$ component, reducing the rate of noise triggers. This method allows an analysis targeting specific areas the parameter space. However, this method still faced issues due to the increase in the rate of noise triggers.

In a third distinct approach, in~\cite{Harry:2016ijz} the authors proposed a search where templates are placed to cover all of the required parameters, including the two masses, six spin components and the inclination of the source~\footnote{If only considering the $l=2$, $|m| = 2$ harmonics in the instantaneous orbital plane orbital phase can still be analytically maximized over. This is not the case if higher-order modes are included.}. Using this method for a single detector, the observed signal will always be consistent with a set of physical parameters, reducing the rate of noise triggers. However, when we attempted to generate a template bank covering a physically meaningful parameter space for neutron-star--black-hole signals, we generated a template bank in excess of $30{,}000{,}000$ templates and it had showed no sign of converging at that point. Filtering gravitational wave data with this many templates and constructing sets of filter waveforms of this size and larger is computationally impractical.

A common issue between these methods to search for precessing signals is that in order to properly model the observed signal, extra parameters must be added to our signal model in order to account for the effects of precession. However, any extra degrees of freedom in the signal model will not only increase the computational cost of the search, but will also increase the rate of noise triggers, even in the case of simple Gaussian noise. When constructing a precessing search we must therefore ensure that any increase in the sensitivity due to an improved signal model is not out-weighed by a relative increase in the noise rate.

\section{\label{sec:harmonics}Precessing waveform harmonic decomposition}

In this section we will review the harmonic decomposition for precessing signals, as introduced in \cite{Fairhurst:2019vut}, before discussing how this formulation can be useful in solving many of the issues reducing the effectiveness of current precessing searches in the next section.

The gravitational waves emitted by a binary can be decomposed into a set of spin-weighted spherical harmonics, given by
\begin{equation}
    \label{eq:spherical_harmonics}
    h_{+} - i h_{\times} = \sum_{l \geq 2} \sum_{-l \leq m \leq l} h_{l,m}(t) Y^{l,m}_{(-2)} (\theta_{\mathrm{JN}}, \varphi_0),
\end{equation}
where $Y^{l,m}_{(-2)}(\theta_{\mathrm{JN}}, \varphi_0)$ are the spin-weighted spherical harmonics with weight $-2$~\cite{Kidder:2007rt}, $h_{l,m}(t)$ are the harmonic components for the binary and $\theta_{\mathrm{JN}}$, $\varphi_0$ are angles giving the direction to the observer in a source-centered coordinate system with the its $z$-axis along $\vect{J}$, as described earlier in section~\ref{sec:precession}.

To calculate the emitted gravitational waves, $h^{\prime}_{+}(t, \theta_{\mathrm{JN}}, \varphi_0)$ and $h^{\prime}_{\times}(t, \theta_{\mathrm{JN}}, \varphi_0)$ for a binary rotated with respect to the original by the Euler angles $(\alpha, \beta, \gamma)$  we can calculate the new harmonic components by performing a rotation
\begin{equation}
    \label{eq:component_rotation}
    h^{\prime}_{l,m}(t) = e^{im\alpha} \!\!\!\!
    \sum_{-l \leq m^{\prime} \leq l} e^{im^{\prime}\gamma} d^l_{m^{\prime}, m}(-\beta) h_{l,m^{\prime}}(t),
\end{equation}
where $d^l_{m, m^{\prime}}$ is the Wigner d-matrix~\cite{Hannam:2013oca}.

An important property of the spin-weighted spherical harmonics is that, under rotation, the modes for a particular value of $l$ will not couple with modes for other $l$ values~\cite{Gualtieri:2008ux}. If we therefore start by restricting ourselves to the dominant $l=2$, $|m| = 2$ mode, after performing a rotation we will have a maximum of five non-zero harmonic components with $l=2$, $-2 \leq m \leq 2$.

In the case of a precessing binary under the adiabatic limit we can therefore calculate the harmonic components, $h^{\text{P}}_{2,m}(t)$, by performing a time-dependent rotation of the harmonic components of a non-precessing binary, $h^{\text{NP}}_{2,m}(t)$~\cite{Schmidt:2012rh}. The first two Euler angles of the rotation are given by $\alpha(t)$ and $\beta(t)$ as defined in Figure~\ref{fig:precession_jz}. The third Euler angle is $\gamma(t)$ which is defined by~\cite{Boyle:2011gg, Fairhurst:2019vut}
\begin{equation}
    \Dot{\gamma} = \Omega_p \cos \beta.
\end{equation}

By combining equations \ref{eq:component_rotation} and \ref{eq:spherical_harmonics} and using the definitions of the spin-weighted spherical harmonics and Wigner d-matrix it is shown in~\cite{Fairhurst:2019vut} that the observed signal for a precessing binary can then be written as
\begin{equation}
\label{eq:precessing_hoft}
\begin{aligned}
    h(t) = \Re \biggl[
    &\frac{A_0(t) e^{2i(\phi_s(t) + \alpha(t))}}
    {(1 + b^2(t))^2} 
    \sum_{k=0}^4 \left( be^{-i \alpha (t)} \right)^k \\
    &\left( F_{+} \mathcal{A}^{+}_k - i F_{\times} \mathcal{A}^{\times}_k \right)
    \biggr],
\end{aligned}
\end{equation}
where the amplitude $A_0(t)$ is proportional to $|h^{\text{NP}}_{2,2}(t)|$. The phase, $\phi_s(t)$, is a combination of the non-precessing waveform's orbital phase and $\gamma$ given by
\begin{equation}
    \phi_s(t) = \phi(t) - \gamma(t),
\end{equation}
the parameter $b(t)$ is defined as
\begin{equation}
    b(t) = \tan (\beta(t) / 2)
\end{equation}
and the constants $\mathcal{A}_k^+$ and $\mathcal{A}_k^{\times}$ are defined as
\begin{align}
    \mathcal{A}_0^+ = \mathcal{A}_4^+ &= \frac{1}{r} \left( \frac{1 + \cos^2 \theta_{\mathrm{JN}}}{2} \right) \nonumber \\
    \mathcal{A}_0^{\times} = - \mathcal{A}_4^{\times} &= \frac{1}{r} \cos \theta_{\mathrm{JN}} \nonumber \\
    \mathcal{A}_1^+ = - \mathcal{A}_3^+ &= \frac{2}{r} \sin \theta_{\mathrm{JN}} \cos \theta_{\mathrm{JN}} \nonumber \\ 
    \mathcal{A}_1^{\times} = \mathcal{A}_3^{\times} &= \frac{2}{r} \sin \theta_{\mathrm{JN}} \nonumber \\ 
    \mathcal{A}_2^+ &= \frac{3}{r} \sin^{2} \theta_{\mathrm{JN}} \nonumber \\ 
    \mathcal{A}_2^{\times} &= 0.
\end{align}

\begin{figure*}
    \centering
    \includegraphics[width=0.98\textwidth]{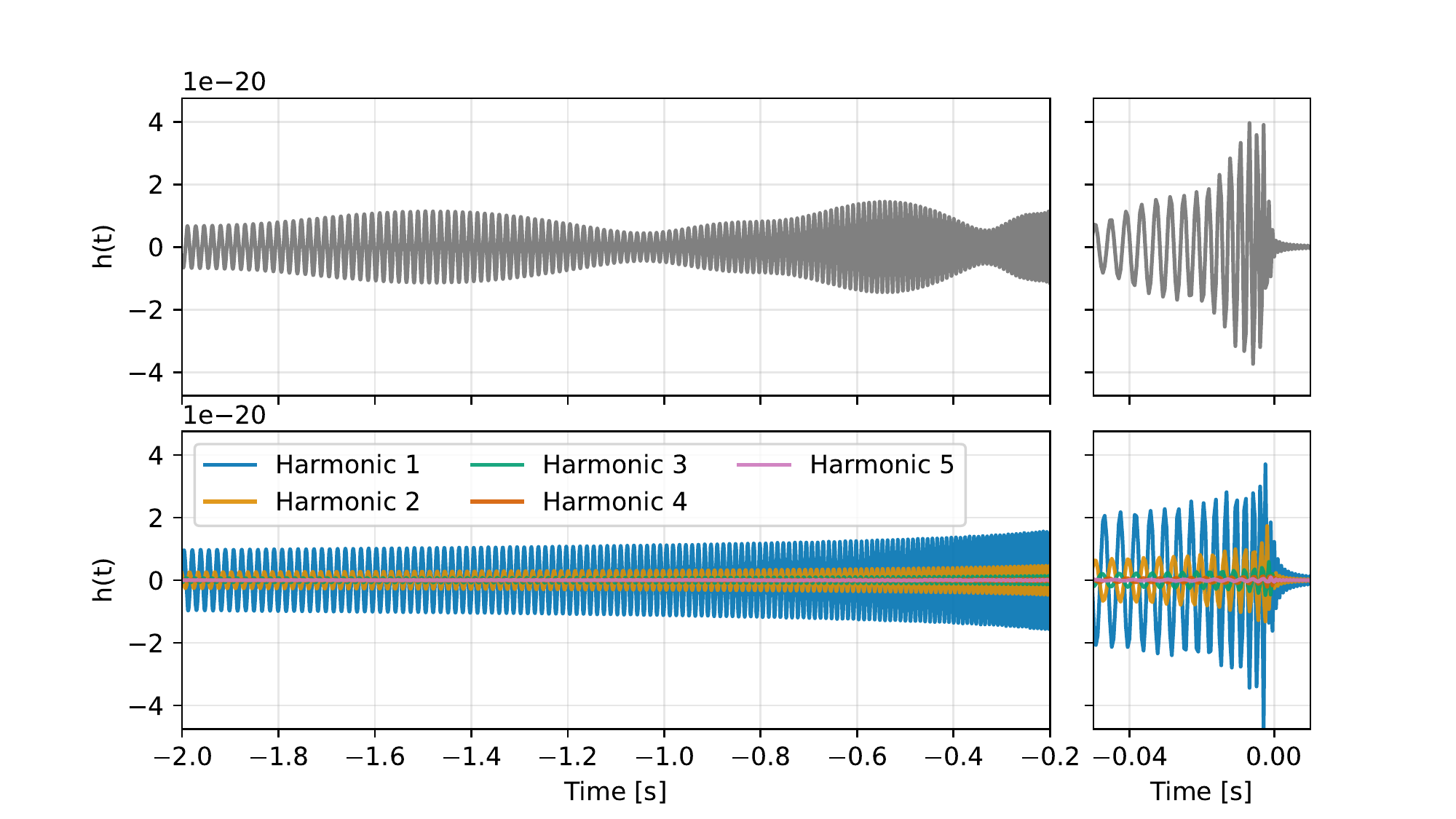}
    \caption{An example of a precessing waveform for a binary with component masses $m_1 = 10 \msun$, $m_2 = 1.5 \msun$, and component spins $\vect{s}_1 = (0.5, 0.5, 0.5)$, $\vect{s}_2 = (0, 0, 0)$. The top panel shows the observed waveform for a signal with $\iota = \pi / 4$ defined at $20\,\mathrm{Hz}$ viewed directly overhead. The bottom panel shows the five harmonics for this template. For both the top and bottom panels, the \emph{Left} plot focuses on the inspiral, and the \emph{Right} plot focuses on the merger and ringdown. }
    \label{fig:harmonics}
\end{figure*}

We can see from equation \ref{eq:precessing_hoft} that the observed signal has five harmonic components forming a power series in $b(t)e^{-i \alpha (t)}$. For each value of $k$ the amplitude is therefore scaled by an extra factor of $b$ and the frequency increases by the precession frequency. Each individual harmonic's amplitude evolves proportionally to the aligned-spin waveform and will therefore not show the characteristic modulation of a precessing signal. The modulation effects are then generated by the interference between the different harmonics. This can be seen in Figure \ref{fig:harmonics}, which shows a precessing signal and the five harmonics generated for the same binary.

Following~\cite{Fairhurst:2019vut} we will factorise out the dependence on the initial orbital phase and precession phase, defining
\begin{equation}
    \hat{\phi} (t) = \phi_s (t) - \phi_0 + \alpha (t) - \alpha_0.
\end{equation}
We can then rewrite the observed signal in the form
\begin{equation}
  h(t) = \Re \sum_{k=0}^4 A_k h_k(t) e^{i \phi_k},
\end{equation}
where
\begin{equation}
  h_k(t) = \frac{A_0(t) b^k(t)}{(1 + b^2(t))^2} e^{i\left(
    2 \hat{\phi} (t) - k(\alpha (t) - \alpha_0) \right)},
\end{equation}
\begin{equation}
    \label{eq:harmonic_dk}
    A_k = \left( (F_{+} \mathcal{A}^{+}_k)^2 + (F_{\times} \mathcal{A}^{\times}_k)^2 \right)^{1/2}
\end{equation}
and
\begin{equation}
    \label{eq:harmonic_phik}
    \phi_k = 2 \phi_0 + (2 - k) \alpha_0 - \tan^{-1}
    \left(
    \frac{F_{\times} \mathcal{A}^{\times}_k}{F_{+} \mathcal{A}^{+}_k}
    \right).
\end{equation}
We can see that the signals' dependence on the extrinsic parameters is contained within the ten constant components of $A_k$ and $\phi_k$. Any change in the sky location or orientation will therefore simply correspond to a change in the overall amplitudes and phases of each harmonic, while the evolution of the amplitude and phase is unchanged.

Using the stationary phase approximation we can express the observed waveform in the frequency domain~\cite{Fairhurst:2019vut} as
\begin{equation}
\label{eq:hk}
  \Tilde{h}(f) = \Re \sum_{k=0}^4 A_k \Tilde{h}_k(f) e^{i \phi_k},
\end{equation}
where
\begin{equation}
    \Tilde{h}_k(f) = \frac{A_0(f) b^k(f)}{(1 + b^2(f))^2} e^{i(2\hat{\phi}(f) - k(\alpha(f) - \alpha_0))}.
\end{equation}

\section{Applying the harmonic decomposition}
\label{sec:search_motivation}

Using the formulation of the observed waveform in the frequency domain as a sum of 5 harmonics described in equation~\ref{eq:hk}, we can formulate a procedure to search
for precessing waveforms using this harmonic construction for the template filter waveforms.
For a given set of intrinsic parameters (masses and spins), we can search using each harmonic individually and maximise over the ten parameters, $D_k$ and $\phi_k$, effectively maximising over the extrinsic parameters of the binary. In order to obtain the templates for each harmonic we linearly combine waveforms generated with different values of the extrinsic parameters $(\Theta, \Phi, \Psi, \theta_{\mathrm{JN}}, \alpha_0, \phi_0)$. A method for achieving this is laid out in \cite{Fairhurst:2019vut}.

However, doing this with all 5 harmonics would be very similar to the method of~\cite{Pan:2003qt} and would suffer the same issues of constraining the $A_k$ and $\phi_k$ values to physically possible combinations.
Nevertheless, the nature of the harmonic decomposition offers a way to solve this problem. As the harmonics form a power-series in the parameter $b$, if we have $b < 1$ then each subsequent harmonic in the series will be weaker than the previous.
Likewise, if $b > 1$ then this will be reversed and the fifth harmonic will be the most significant.
In~\cite{Fairhurst:2019vut} the authors show that for the majority of binaries in the sensitive range of current detectors the average value of $b$ is below 0.4, in this case each subsequent harmonic after the first will be less significant than the previous when modelling the precessing signal.
This provides a way to solve the problems faced in previous precessing searches.
If precessing signals can be reliably modelled using less than five harmonics then we can use a smaller number of harmonics in the search.
This will in turn reduce the freedom of the model to match with noise in the data, reducing the noise rate.
In~\cite{Fairhurst:2019vut}, it is suggested to perform a search using only the $k=0,1$ harmonics.
Here, we will investigate the best number of harmonics to be used in order to maximise the sensitivity of the search to precessing signals.

\section{\label{sec:precessing_search}Search setup}

In this section we will describe our implementation of the harmonic decomposition from~\cite{Fairhurst:2019vut} to perform a search for precessing binaries.

For the purpose of this work we will focus on the observation of neutron-star--black-holes as these systems are ones where the effects of precession are most observable~\cite{Brown:2012qf, Harry:2013tca, Capano:2016dsf, Green:2020ptm}. Specifically, we will search for signals with black hole masses in the range $[5, 20] \msun$ and neutron star masses in the range $[1.2, 1.7] \msun$, with maximum spin magnitudes on each component of $0.99$. 

As a starting point we will use a two detector network consisting of the LIGO Hanford and LIGO Livingston detectors~\cite{LIGOScientific:2014pky}, but this method could be extended to a larger network in the future.

\subsection{Waveform model}

To model the precessing signals, we use the \texttt{IMRPhenomXP} waveform model~\cite{Pratten:2020fqn, Garcia-Quiros:2020qpx}. This two-spin model constructs precessing signals by taking an underlying aligned-spin waveform~\cite{Pratten:2020fqn} and performing a time-dependent rotation to model the precession effects~\cite{Schmidt:2012rh}.

Previously, it has been shown that the four in-plane spin degrees of freedom can be mapped to a single parameter, which captures the dominant precession effects. This effective spin precession parameter, $\chi_p$, is defined as~\cite{Schmidt:2014iyl},
\begin{equation}
    \chi_p = \frac{1}{A_1 m_1^2} \max (A_1 S_{1\bot}, A_2 S_{2\bot})
\end{equation}
where
\begin{equation}
    A_1 = 2 + \frac{3 m_2}{2 m_1}
    , \quad \quad
    A_2 = 2 + \frac{3 m_1}{2 m_2}
\end{equation}
and $S_{1\bot}$, $S_{2\bot}$ are the in-plane component spins perpendicular to the orbital angular momentum. We note that other metrics have also been proposed~\cite{Fairhurst:2019vut, Fairhurst:2019srr, Gerosa:2020aiw, Thomas:2020uqj}.

\subsection{Matched filter}

First let us consider a single detector, for a signal with known intrinsic parameters. The log likelihood-ratio for a known signal in Gaussian noise is given by
\begin{equation}
    \lambda(h) = (h|s) - \frac{1}{2} (h|h).
\end{equation}
where $(a|b)$ represents the commonly used noise-weighted inner product
\begin{equation}
    ( a|b ) = 4 \int^{+ \infty}_{0} \frac{\Tilde{a}^{*}(f) \Tilde{b}(f)}{S_n(f)} df.
\end{equation}
As shown in~\cite{Allen:2005fk,Babak:2012zx} for a signal with an unknown amplitude and phase we can maximise the log likelihood-ratio by using the phase-maximised matched-filter
\begin{equation}
    \rho^2 = \frac{|( h|s )|^2}{(h|h)}.
\end{equation}
If the 5 harmonics are independent we can matched-filter over each of them independently while maximizing over the phase ($\phi_k$) and amplitude ($A_k$). However, the harmonics are not guaranteed to be orthogonal to one another. This will introduce covariance between the matched-filter outputs produced by each harmonic, making the maximisation of the log likelihood ratio, or signal-to-noise ratio, more complicated to compute. In order to simplify this calculation we will first ensure that the harmonics are orthogonal and normalised such that
\begin{equation}
    \mathcal{M}_{kl} = \delta_{kl}
    \quad \text{where} \quad
    \mathcal{M}_{kl} = | \langle h_k|h_l \rangle |.
\end{equation}
In order to diagonalize the matrix $\mathcal{M}_{kl}$, while maintaining the natural hierarchy of the harmonics we use the Gram-Schmidt process, where the first orthogonal harmonic is given by $\Tilde{h}_{0\bot} = \Tilde{h}_{0}$ and each orthogonal harmonic for $k > 0$ is given by
\begin{equation}
    \Tilde{h}_{k\bot} = \Tilde{h}_{k} - \sum_{l=0}^{k-1} \langle h_{l}|h_{k} \rangle \Tilde{h}_{l}
\end{equation}
Finally, the orthonormalized harmonics are given by
\begin{equation}
    \Hat{h}_{k\bot} = \frac{\Tilde{h}_{k\bot}}{(h_{k\bot}|h_{k\bot})}
\end{equation}
After completing this step we can maximise the log likelihood-ratio by summing the phase-maximised signal-to-noise ratios for each harmonic in quadrature
\begin{equation}
    \rho_h^2 = \max_{A_k, \phi_k} \left[ \rho^2 \right] = \sum^N_{k=1} \rho^2_k
\end{equation}
where $\rho_h$ is the total signal-to-noise ratio and $N$ is the number of harmonics we choose to use, the choice of which we will discuss shortly. This gives a simple method to maximise the signal-to-noise ratio over the sky location and orientation, capturing the full (if $N=5$) signal-to-noise ratio of the precessing signal. The phase-maximised signal-to-noise ratio for each harmonic, $\rho^2_k$, is the sum of two independent Gaussian random variables, each with a mean of 0 and variance of 1, in the absence of a signal~\cite{Allen:2005fk}. The value of $\rho^2_h$ in the absence of a signal will therefore be the sum of $2N$ Gaussian random variables following a $\chi^2$-distribution with 2N degrees of freedom. 

The values of $A_k$ and $\phi_k$ are not independent. The ten components of $A_k$ and $\phi_k$ are dependent on a total of seven parameters ($r$, $\Theta$, $\Phi$, $\Psi$, $\phi_0$, $\theta_{\mathrm{JN}}$, $\alpha_0$). When considering a single detector, the three detector response angles and the distance $r$ enter as two independent quantities; an overall scaling factor and the ratio between the response factors $F_+$ and $F_{\times}$. This means that the ten components of $A_k$ and $\phi_k$ are functions of five independent parameters and there must therefore be constraints on the allowed values of $A_k$ and $\phi_k$. Maximising over all ten components of $A_k$ and $\phi_k$ independently will allow for many unphysical combinations, increasing the expected noise for the matched filter.

However, as suggested in~\cite{Fairhurst:2019vut}, due to the natural hierarchy of the harmonics it may not be necessary to use the full set of five harmonics for each template. When $b < 1$ each harmonic after the first will be weaker by a factor of $b$ and will contribute less signal-to-noise ratio to the observed signal. For different sky positions and orientations the relative amplitudes of the harmonics will also change as we can see in equation \ref{eq:harmonic_dk}. In some cases, even when $b$ is small, harmonics with larger $k$ values will be able to contribute significantly to the signal-to-noise ratio, however, for most binary configurations this will be rare.

\begin{figure}
    \centering
    \includegraphics[width=0.45\textwidth]{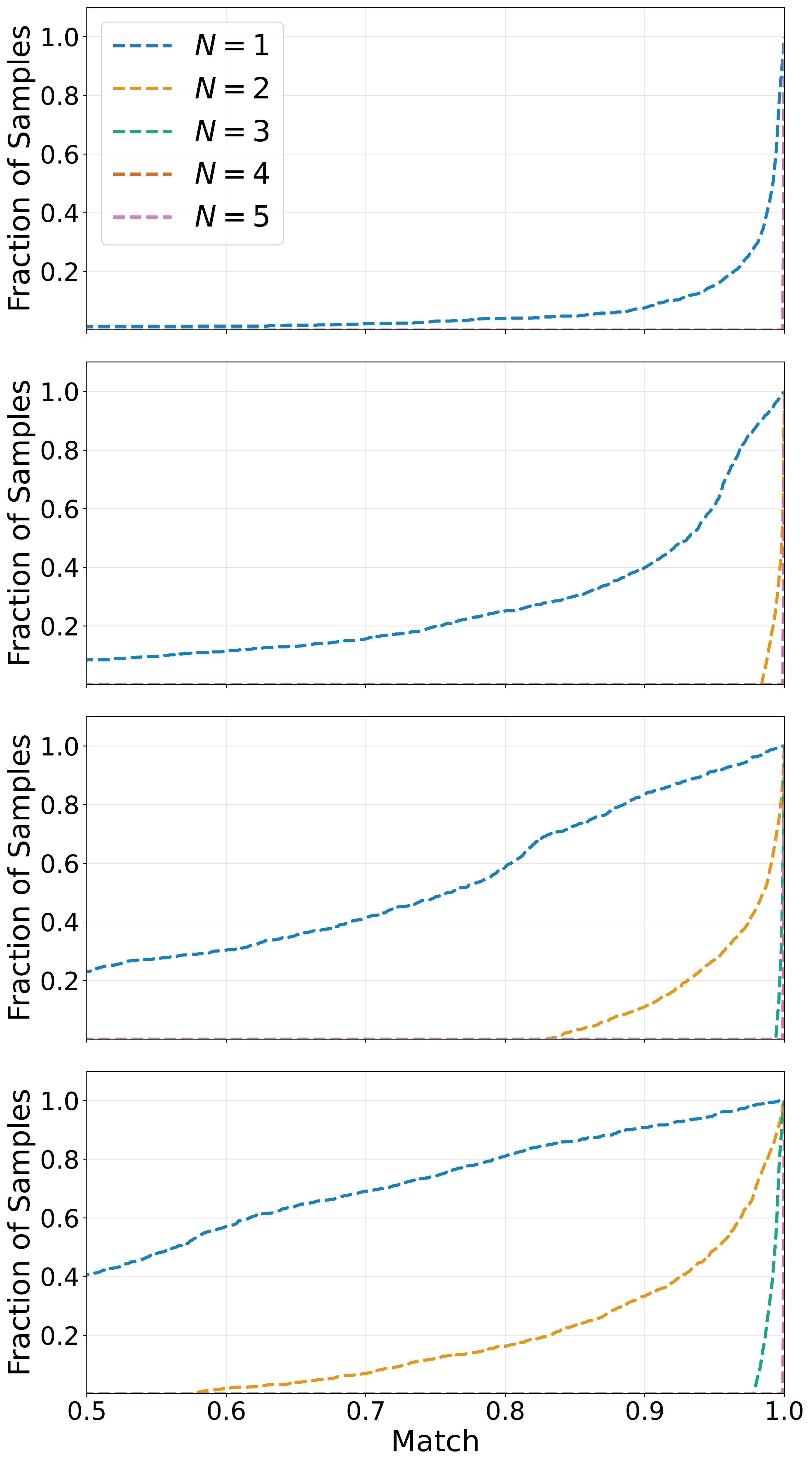}
    \caption{Distribution of matches for four binaries with component masses $m_1 = 10 \msun$, $m_2 = 1.5 \msun$, component spins parallel to the orbital angular momentum $s_{1\parallel} = 0.3$, $s_{2\parallel} = 0.3$, and a precessing spins (from top to bottom) of $\chi_p = 0.1$, $\chi_p = 0.3$, $\chi_p = 0.6$, $\chi_p = 0.9$. Each sample is drawn with a random sky position and orientation. The match is calculated using a template with matching intrinsic parameters using different numbers of harmonics $N$.}
    \label{fig:1D_overlaps}
\end{figure}

As an example, in figure \ref{fig:1D_overlaps} we show the distribution of matches for four binaries with $m_1 = 10 \msun$, $m_2 = 1.5 \msun$, component spins parallel to the orbital angular momentum $s_{1\parallel} = 0.3$, $s_{2\parallel} = 0.3$, and perpendicual spin components of $\chi_p = 0.1$, $\chi_p = 0.3$, $\chi_p = 0.6$, $\chi_p = 0.9$ respectively. We randomly generate a large number of systems with isotropic sky positions and orientations. We calculate the match using a template with matching intrinsic parameters using different numbers of harmonics. We see that in the case where the in-plane spin is the smallest, $\chi_p = 0.1$, a single harmonic is able to achieve a match above $0.97$ for $\sim 79\%$ of the tested sky positions and orientation. When using two harmonics this threshold is passed for $100\%$ of the samples. In this case using a third harmonic would make no difference to the match with signals in the data. When the in-plane spin is increased and precession effects become stronger the match using the first two harmonics is reduced. In the most extreme case considered in this example, $\chi_p = 0.9$, we see that using two harmonics achieves a match above $0.97$ for only $\sim 40\%$ of the samples. However, if using three harmonics we find matches larger than $0.97$ for all points tested.

\begin{figure}
    \centering
    \includegraphics[width=0.48\textwidth]{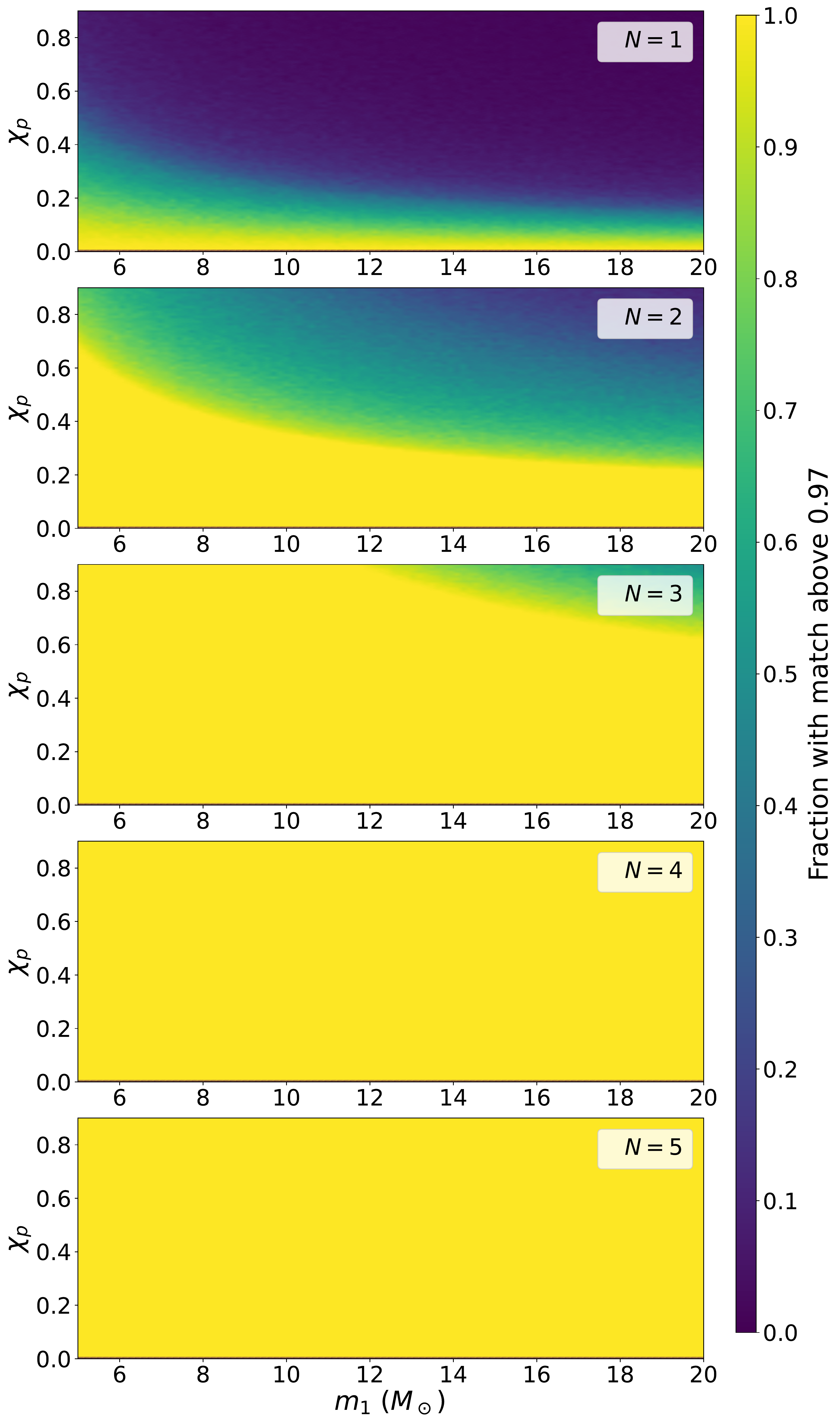}
    \caption{The fraction of sky positions and orientations for which the match is over $0.97$, using (from top to bottom) one harmonic, two harmonics, three harmonics, four harmonic and five harmonics. Matches are calculated for binaries with a fixed secondary mass, $m_2 = 1.5 \msun$, and component spins parallel to the orbital angular momentum $s_{1\parallel} = 0.3$, $s_{2\parallel} = 0.3$, while the primary mass $m_1$ and in-plane spin, $\chi_p$, are varied. Matches are calculated using templates with matching intrinsic parameters.}
    \label{fig:match_map}
\end{figure}

In figure \ref{fig:match_map} we consider systems where we fix $m2=1.5 \msun$ and component spins parallel to the orbital angular momentum $s_{1\parallel} = 0.3$, $s_{2\parallel} = 0.3$ but allow the larger mass, $m1$ and the perpendicular spin $\chi_p$ to vary. As in figure~\ref{fig:1D_overlaps} we then randomly generate a large set of points with an isotropic distribution of sky location and orientation and compute the fraction of signals with matches greater than $0.97$ when filtering with a varying number of harmonics. We see that there is only a small region of the parameter space where $\chi_p \approx 0$ that a single harmonic is able to reliably recover the majority of the signal-to-noise ratio. A much larger region of the parameter space is covered by increasing to two harmonics, while using three harmonics has average matches above $0.97$ for all but a small region with high mass ratios and large $\chi_p$. Extending to four or five harmonics achieves an average match of $> 0.97$ for all configurations considered in this example. We do release all the code used to make these plots in~\cite{DATA_RELEASE} so that an interested reader can easily generate these figures of merit for different mass and/or spin configurations.

These results motivate a scenario where $N$, the number of harmonics we filter with for any template, is dependent on the importance of the sub-dominant harmonics for that template. In the search we demonstrate here we will attempt to filter with the smallest number of harmonics for any template while still maintaining a match above $0.97$ for the majority of binary configurations of that template. In this case, we will still have examples where there is unphysical freedom (i.e. where $N$ is large for a given template), and there are some combinations of $A_k$ and $\phi_k$, which while physical, may be statistically very unlikely. It would be best to impose suitable priors on $A_k$ and $\phi_k$ such that that we could include additional harmonics without penalizing the search. Doing this in practice is complicated, but we discuss this more, and present an approximate solution, later in this work.

\subsection{\label{sec:harmonic_bank}Template bank generation}

In order to generate our set of filter waveforms (or template bank) for this search we will use stochastic template bank generation~\cite{Harry:2009ea, Ajith:2012mn}.

In the case of a search with aligned-spin filter waveforms stochastic template bank generation works as follows. Starting with an empty template bank a random point is chosen from within the target parameter space and the corresponding signal, $h_{\text{prop}}$, is generated. For each template, $h_i$, within the current template bank the match, $m(h_i, h_{\text{prop}})$ is calculated and the maximum match across the template bank is defined as the fitting factor
\begin{equation}
    \text{FF}(h_{\text{prop}}) = \max_{i} \left[ m(h_i, h_{\text{prop}}) \right]
\end{equation}
If the fitting factor is below a given threshold, usually 0.97, then the proposed template $h_{\text{prop}}$ is not covered sufficiently by the current template bank and $h_{\text{prop}}$ is added to the bank. This process is repeated iteratively until a set limit for the size or coverage of the template bank is reached.

In the case of aligned-spin filter waveforms the match is calculated using the phase-maximised matched-filter. In that case we do not need to maximise over the phase and time of the proposal $h_{\text{prop}}$, as the effect it would have on the match would be equivalent to a phase or time shift in the template $h_i$. The match therefore does not depend on the sky position or orientation.

In the case of a precessing signal if we maximise over the values of $A_k$ and $\phi_k$ for the bank template, $h_i$, the match will still depend on the values of $A_k$ and $\phi_k$ of the proposal point $h_{\text{prop}}$. Therefore when proposing a new point we choose a specific sky location and orientation giving us specific values of $A_k$ and $\phi_k$ for $h_{\text{prop}}$. We then calculate the match as
\begin{equation}
    \label{eq:harmonic_m}
    m(h_i, h_{\text{prop}}) = \max_{t_i, t_{\text{prop}}} \left[ \sum^5_{k=1} \frac{|\langle\hat{h}_{k \bot,i} | h_{\text{prop}} \rangle \rangle|^2}{(h_{\text{prop}}|h_{\text{prop}})} \right]^{1/2},
\end{equation}
noting that we explicitly include all 5 components during template bank generation.
The match is then maximised across the template bank to get the fitting factor. If the fitting factor is less than the threshold the point is then added to the bank, discarding the sky location and orientation parameters.
We do note that this process may lead to a high density of templates in some regions of parameter space where for specific values of the binary orientation the waveform can vary significantly with small changes in other parameters.

It is also very difficult to assess when the template bank has fully converged, as we have not managed to generate any template bank which achieves a fitting factor $> 0.97$ in all areas of parameter space. We instead empirically evaluated the coverage of the bank as it grew larger and stopped at 360,000 templates. At this point, choices we will consider later regarding the maximum number of harmonics we use in the matched-filtering limit the coverage of the bank more than the small loss shown here when allowing up to 5 harmonics to be used. Using this process we generate a bank of 358{,}866 templates. While testing this template bank, we observed a significant number of templates included unphysical features and didn't follow the expected harmonic heirarchy. We discuss this more in Appendix~\ref{sec:app_phenomx_inconsistencies} but these issues resulted in us removing 3{,}706 templates from the template bank, resuling in a final bank of 355{,}160.
We evaluate the performance of this template bank in the next subsection.

\subsection{Selecting the number of harmonics for each template}

After creating the template bank we then want to choose the number of harmonics, $N$, to use for each template. For each template we generate a set of $n$ samples with randomly drawn sky locations and orientations, $\vect{\Omega}$, and matching masses and spins. We then calculate an effective match~\cite{Buonanno:2002fy}
\begin{equation}
    \label{eq:harmonic_meff}
    m_\text{eff}(h_i) = \left[ \frac{\sum_{i=0}^n m(h_i, h_i(\vect{\Omega}))^3 \rho_\text{opt}^3(h_i)}{\sum_{i=0}^n \rho_\text{opt}^3(h_i)} \right]^{1/3}
\end{equation}
where
\begin{equation}
    \rho_\text{opt}^2(h) = \left[ \sum^4_{k=0} |\langle \hat{h}_{k \bot} | h \rangle \rangle|^2 \right]^{1/2}
\end{equation}
The effective match weights each match by the observable volume of that signal. Maximising the effective match therefore favours the sky positions and orientations that we are most likely to observe.

We start by calculating this value using only one harmonic in the match calculation and increase the number of harmonics until the effective match is greater than $0.97$, recording the number of harmonics used as $N$ for that template. This process is repeated for the full bank in order to minimise the number of harmonics that are used, reducing the expected noise background.

The exact threshold used for $m_\text{eff}(h_i)$ when choosing $N$ could be tuned in the future in order to make the optimal trade-off between an increase in the template banks' sensitivity to precessing signals and the increase in noise due to larger values of $N$. For this work we will continue with $0.97$.

In the future it may also be preferable to choose the value of $N$ during the construction of the template bank so that the choice of $N$ is taken into account when testing matches for templates with different intrinsic parameters. This could be achieved by calculating the effective match between each template and the proposal within the template bank generation loop. However, this would require some optimisation to be computationally feasible and we leave this for future work. In the data release accompanying this paper we provide some additional thoughts, and a partial implementation, of how to achieve this.

\begin{figure*}
    \centering
    \includegraphics[width=0.98\textwidth]{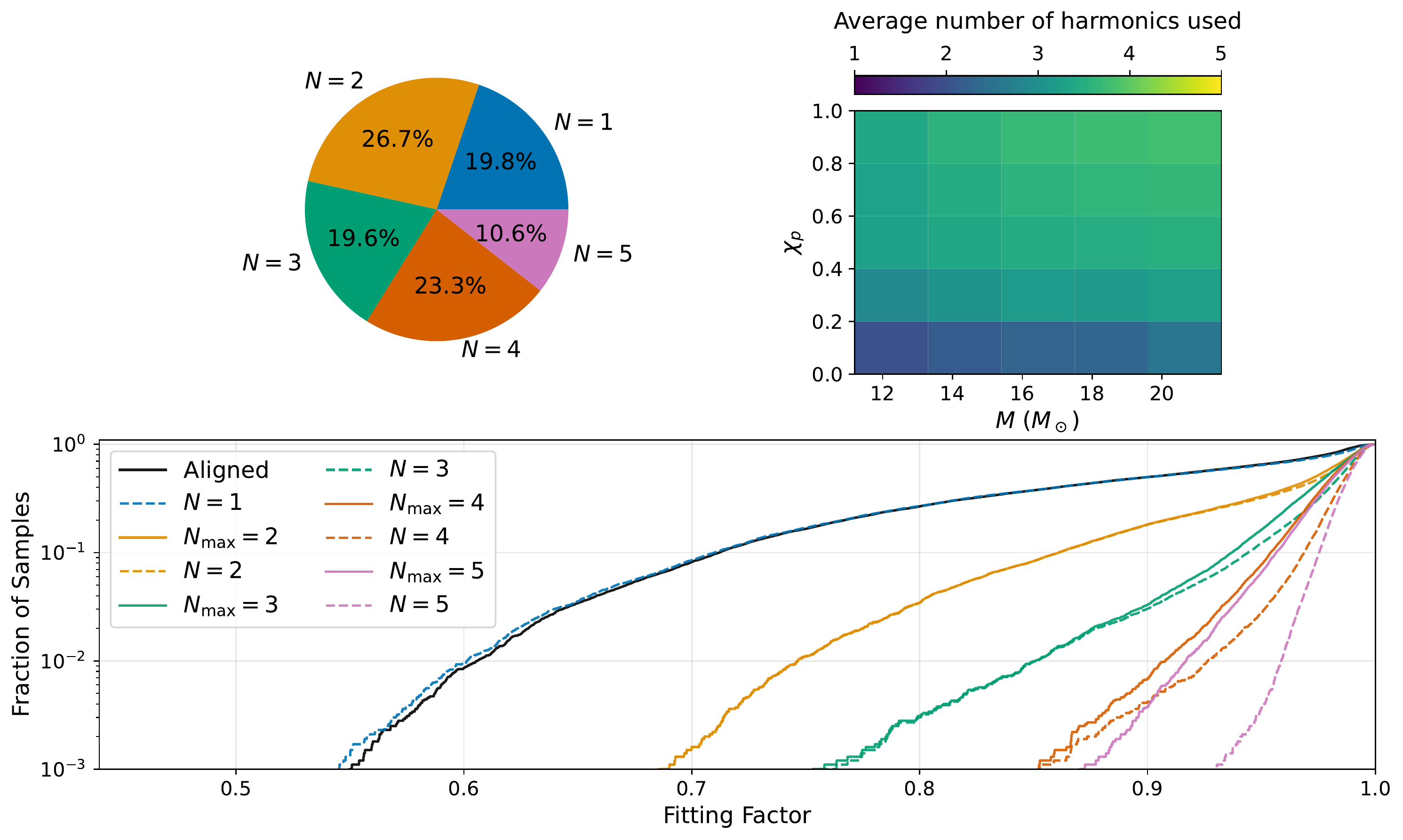}
    \caption{\emph{Top row}: The number of harmonics selected for the templates in the harmonic template bank. The left panel shows the percentage of the template bank that use one, two or three harmonics. The right panel shows the average number of harmonics selected within a set of bins over the total mass and in-plane spin parameter. \emph{Bottom row}: The number of simulated signals with fitting factors above a given value. We generate a set of simulated signals within the target parameters and calculate the fitting factors for both the aligned-spin and precessing template banks. We show six different configurations of the template bank. The black line shows the fitting factor when using the aligned-spin template bank. Each other colour shows the fitting factor using a different maximum number of harmonics. The solid lines show the fitting factor when using the value $N$ calculated for each template in the bank, while the dashed lines show the fitting factor if we used the maximum number of harmonics for each template. }
    \label{fig:num_comps}
\end{figure*}

In the top row of Fig. \ref{fig:num_comps} we see the number of templates using 1, 2, 3, 4 or 5 harmonics. There is a reasonably even split between the 5 possibilities, perhaps surprisingly, as many as 10\% of the templates require 5 harmonics to achieve our figure of merit. As expected, the average number of harmonics selected increases as the in-plane spin and mass (and therefore mass-ratio) increase, where the effects of precession will be strongest (see e.g. Ref.~\cite{Green:2020ptm}).

In the bottom row of Figure ~\ref{fig:num_comps} we demonstrate the effectiveness of our precessing template bank by computing fitting factors using a varying number of harmonics (either always using N harmonics, or limiting templates to no more than N harmonics). The simulated signals are drawn with a uniform distribution of component masses and spins within the parameter space considered. Spin, sky location and orientation angles are drawn isotropically. This is also compared to the performance with an aligned-spin template bank, which contains 118{,}837 templates covering aligned-spin signals with a minimum match of 0.97 within the target region. The aligned-spin template bank was generated with the \texttt{IMRPhenomXAS} \cite{Pratten:2020fqn} waveform model.
When using all five harmonics, the precessing template bank performs very well, with the vast majority of fitting factors larger than 0.95. When restricting to two harmonics there is a noticeable tail of low fitting factors---as low as 0.7---but it is a significant improvement over the aligned-spin template bank, or only using one harmonic. Three and four harmonics respectively all provide additional significant improvement. Allowing a dynamic choice of the number of harmonics for each template results in a template bank where over 99\% the templates are covered with a fitting factor larger than 0.9. This is the first demonstration of such a high fitting factor for a broad range of neutron-star black systems when considered second-generation gravitational-wave interferometer sensitivity curves.

\section{\label{sec:harmonic_coinc}Coincident search}

We will now test the effect that the new template bank has on the sensitivity of a coincident modelled search. We use a stretch of $\sim 8$ days of data in the first half of the third LIGO-Virgo observing run from 14:42:36 21/05/2019 UTC to 10:38:20 29/05/2019 UTC. This data is available from GWOSC \cite{Vallisneri:2014vxa, LIGOScientific:2019lzm}.

We search this data 3 times. First, using our aligned-spin template bank, to set a baseline. Second, using the precessing template bank, limited to a maximum of 2 harmonics. Third, using the precessing template bank, limited to a maximum of 3 harmonics. Although using up to 5 harmonics would give the best results, as motivated in the previous section, it increases the computational cost; we expect the computational cost to scale with the number of harmonics, and the increase in the template bank. When limiting the bank to a maximum of 2 harmonics, 3 harmonics and 5 harmonics, we expect the computational cost to increase by $\sim 5\times$, $\sim 7\times$ and $\sim 8\times$ respectively. We therefore limit our search to a maximum of 3 harmonics to a) reflect the most likely configuration for online and offline analyses and b) provide the best compromise between computational cost and finding precessing gravitational-wave signals. Additionally, the ranking statistic that we will introduce in section~\ref{ssec:ranking_statistic} for our precessing search
has not been defined for more than 3 harmonics. We will discuss
extending the search at the end of this work, but focus here
on demonstrating that this method is viable for finding
precessing compact binary objects.

In order to evaluate the sensitivity of the search a set of simulated signals are added into the data and evaluated in the same way as the real data. The masses are drawn from a log-normal distribution over the target parameter space and the spin magnitudes are drawn uniformly with the larger body having spin magnitudes up to 0.99 and the second body up to 0.05. The spin orientations, sky location and orientation angles are drawn isotropically. The signals are then generated using \texttt{IMRPhenomXP} \cite{Pratten:2020ceb}. For all simulated signals the distance is drawn uniformly in chirp-mass weighted distance with a distance chosen uniformly in $[1, 100]$ Mpc and then multiplied by $\mathcal{M}^{5/6} / 1.2187$ (the chirp mass of a double neutron star with both components having a mass of 1.4).

\subsection{Signal-consistency tests}

There are many instances of non-Gaussian noise within the detector data. In order to mitigate their effects the aligned-spin PyCBC search uses two $\chi^2$ tests to define a new single-detector ranking statistic which down-weights times where the detector noise is non-Gaussian, assuming the presence of a signal \cite{Davies:2020tsx}.

First we apply the $\chi^2$ test described in \cite{Allen:2004gu} which tests the distribution of power in different frequency bins. However, the power of the observed waveform as a function of frequency will change with the sky location and orientation due to the change in the relative amplitudes of the different harmonics. We will therefore reconstruct the combination of harmonics which maximised the signal-to-noise ratio for a particular trigger and use this signal to re-filter the data and calculate the value of the $\chi^2$ test. This method was previously used when developing a search for compact binary coalescences with higher harmonics in \cite{Harry:2017weg} and implemented in a search in \cite{Chandra:2022ixv}.

In order to reconstruct the signal which maximised the signal-to-noise ratio we multiply the orthogonalised harmonics, $h_{k \bot}$, for a particular trigger by their complex signal-to-noise ratio and sum over the $N$ harmonics
\begin{equation}
    h = \sum_{k=0}^N \langle \hat{h}_{k \bot}|s \rangle \hat{h}_{k \bot}.
\end{equation}
This signal can then be used to calculate the boundaries of the frequency bins such that the signal-to-noise ratio is evenly distributed between them and the reduced $\chi^2$ value calculated as
\begin{equation}
    \chi_r^2 = \frac{p}{2p - 2} \sum^{p}_{i=1} \left( \frac{\rho}{p} - \rho_{bin,i} \right)^2.
\end{equation}
where $p$ is the number of frequency bins to be used and $\rho_{bin,i}$ is the signal-to-noise ratio in bin $i$. The signal-to-noise ratio is then down-weighted for large values of $\chi_r^2$ calculating a new re-weighted signal-to-noise ratio
\begin{equation}\label{eq:BAreweight}
    \Tilde{\rho} = 
    \begin{cases}
    \rho, & \text{if $\chi_r^2 \leq 1$, } \\
    \rho \left[ \left(1 + (\chi_r^2)^3\right) /2 \right]^{-\frac{1}{6}}, & \text{if $\chi_r^2 > 1$.}
    \end{cases}
\end{equation}

If the waveform model used is not able to capture the full power of the observed signal then there will be residual power in the data due to this mismatch, increasing the value of the $\chi^2$ test and reducing the significance of the signal. An improvement in the match from an improved template bank does not only increase the signal-to-noise ratio of the observed signal, but increases the robustness of the $\chi^2$ test to signals in the data.

\begin{figure}
    \centering
    \includegraphics[width=0.48\textwidth]{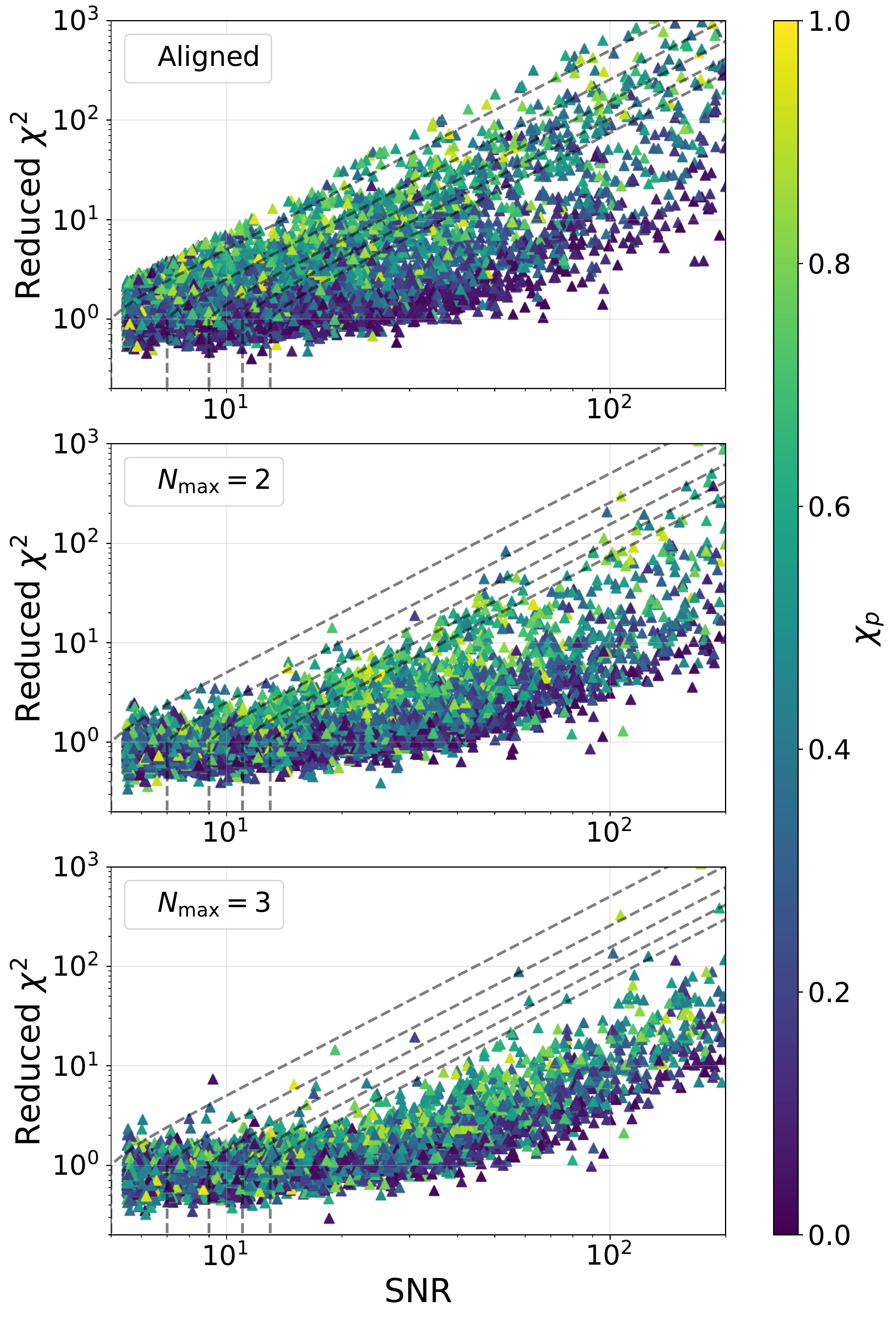}
    \caption{Distribution of a set of simulated precessing signals in the signal-to-noise ratio-$\chi_r^2$ plane. The dashed lines illustrate contours of constant re-weighted signal-to-noise ratio. The top panel shows results for the aligned-spin template bank. The middle and bottom panels show results for the harmonic template bank with a maximum of two and three harmonics respectively.}
    \label{fig:harmonic_chisq}
\end{figure}

In figure \ref{fig:harmonic_chisq} we show the observed signal-to-noise ratios and $\chi_r^2$ values for a number of the simulated precessing signals in the data. We show results for the aligned-spin template bank and the harmonic template bank with a maximum of two and three harmonics. We see that the $\chi_r^2$ value is reduced for simulated signals when using the harmonic template bank, particularly in the case of large in-plane spins, this is due to the increase match with the simulated signals. We do still observe increased values of $\chi_r^2$ for large in-plane spins in the case of the harmonic template banks; this would most likely be improved by using a larger template bank with better coverage of the parameter space and/or by using more harmonics.

We then apply the sine-Gaussian $\chi^2$ test described in \cite{Nitz:2017lco}, which tests for excess power at frequencies above the final frequency of the search template. This step is performed using the same method as described in \cite{Nitz:2017lco}, re-weighting the signal-to-noise ratio again to produce the final single-detector ranking statistic.

\begin{figure}
    \centering
    \includegraphics[width=0.45\textwidth]{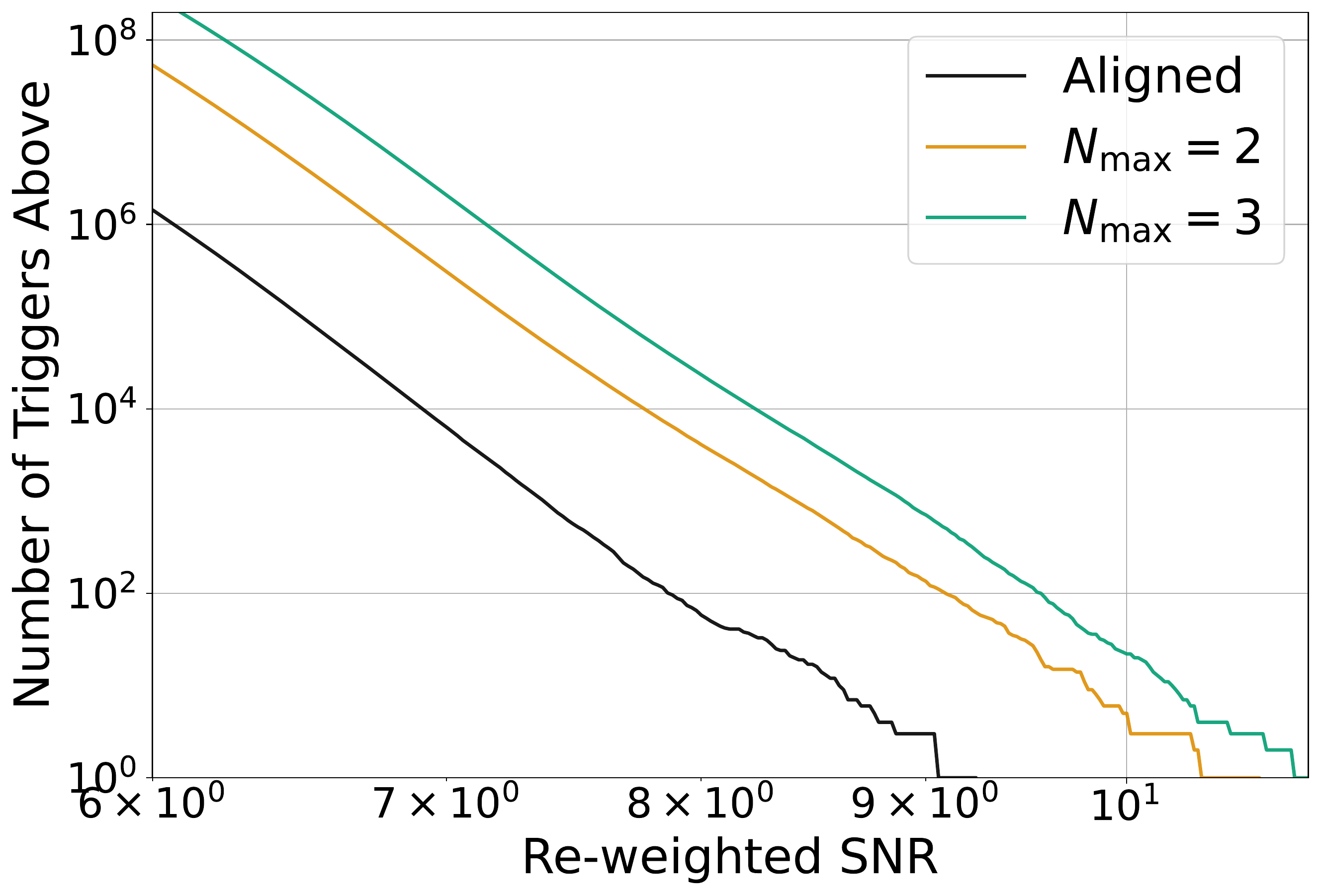}
    \caption{Distribution of the single detector ranking statistic for the LIGO Livingstone detector, showing the number of trigger with re-weighted signal-to-noise ratios above a given value. The black line shows the distribution for the aligned-spin search, while the orange and green lines show the distributions for the harmonic search using a maximum of two and three harmonics respectively.}
    \label{fig:harmonic_singles}
\end{figure}

In figure \ref{fig:harmonic_singles} we show the distribution of re-weighted signal-to-noise ratio for the three template banks. As expected when using the harmonic template bank the noise rate is increased with respect to the aligned-spin template bank. When increasing to a maximum of three harmonics we see a further increase in the noise rate. In order to achieve the same false-alarm rate with a single detector we will therefore need to observe larger signal-to-noise ratios. For example, in order to achieve the same false-alarm rate as a signal-to-noise ratio $7$ trigger in the aligned-spin search we would need to observe a signal-to-noise ratio of $\sim 7.9$ in the case of the two harmonic template bank and a signal-to-noise ratio of $\sim 8.4$ in the three harmonic template bank.

\subsection{Coincident triggers}

After computing the re-weighted signal-to-noise ratio for each detector and identifying a set of single detector triggers, those triggers are compared across the detector network. Triggers are accepted as coincident triggers if they are generated by the same template in the bank and fall within a set time window of each other which is equal to the light travel time between the detectors, plus a small value to account for timing errors.

After generating a set of coincident triggers we must evaluate the significance of each trigger. As a first test we will use the quadrature sum of the single detector trigger's signal-to-noise ratios.

After the coincident signal-to-noise ratio is calculated we perform time-slides of the data in order to generate a set of background coincidences~\cite{Babak:2012zx}. Each coincident trigger is then assigned a false-alarm rate based on this background estimate.

We compare the sensitivity of our three searches using this statistic. We apply a threshold on the false-alarm rate of 1 per 100 years to our set of injections and compute the detection efficiency using fifty distance bins. The volume contained in each distance bin is then multiplied by its detection efficiency and the results are summed across all bins to calculate the sensitive volume.

\begin{figure}
    \centering
    \includegraphics[width=0.45\textwidth]{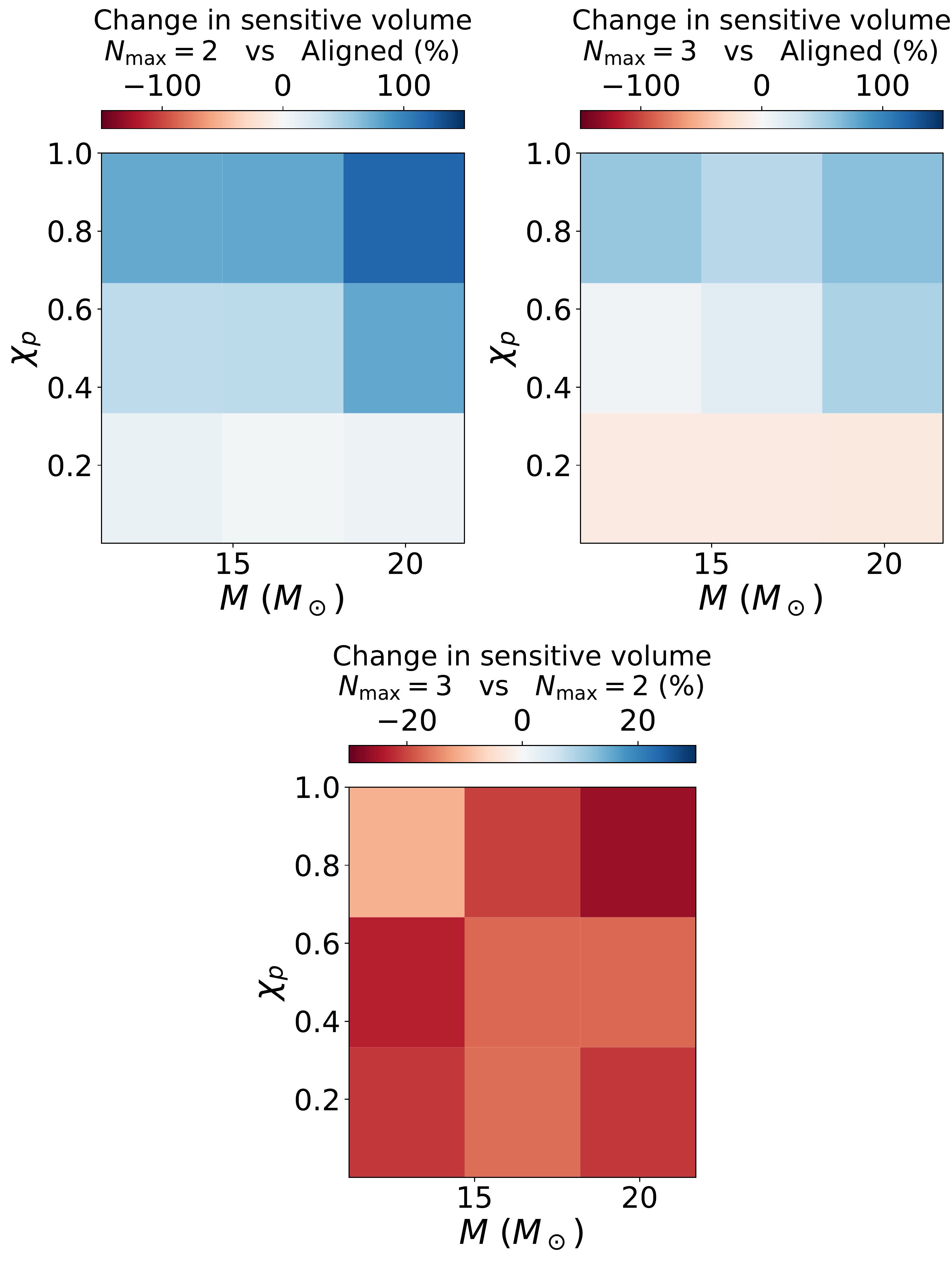}
    \caption{The change in sensitive volume of the search due to the harmonic template bank when using the coincident signal-to-noise ratio. The top left and top right panels show the change in sensitive volume from the aligned-spin template bank to the harmonic template bank with a maximum of two and three harmonics respectively. The bottom panel shows the change in the sensitive volume between the harmonic template bank with a maximum of two and three harmonics respectively.}
    \label{fig:coinc_vt}
\end{figure}

The change in sensitive volume as a function of the total mass and in-plane spin are shown in figure \ref{fig:coinc_vt}. When using a maximum of two harmonics we see a large improvement in the sensitivity of the search for signals with large in-plane spins, increasing the sensitive volume by more than $\sim 50\%$ for signals with $\chi_p > 0.67$. For small values of the in-plane spin we see almost no change in the sensitive volume of the search with respect to the aligned-spin search. This region is expected to show only weak precession effects so we do not expect to see a large improvement in the recovered signal-to-noise ratio, while the increased rate of noise triggers when compared to the aligned-spin case will cause an increase in the false-alarm rate for a fixed signal-to-noise ratio.

When using a maximum of three harmonics we see still see an improvement compared to the aligned-spin search for large in-plane spins, but we now see a decrease in the sensitivity for small in-plane spins. This is due to the increase in the noise rate being more significant than the signal-to-noise ratio gained in this region of the parameter space. We see that across the full parameter space the harmonic search using a maximum of three harmonics is less sensitive than the harmonic search using a maximum of two, again due to the increase in the rate of noise out-weighing any increase in signal-to-noise ratio. In order to include more than two harmonics we will therefore need to improve the ranking statistic being used to reduce the rate of coincident noise triggers.

\subsection{A ranking statistic for our new precessing search}
\label{ssec:ranking_statistic}

The coincident reweighted signal-to-noise ratio statistic is not an optimal ranking statistic. It discards a lot of useful information about the coincident triggers, which could be used to differentiate between signals and noise, such as the relative signal-to-noise ratios and phases of each harmonic for each detector. For each coincident trigger we have a list of properties
\begin{equation}
    \vect{\kappa} = (\left[ \rho_{k,d}, \phi_{\text{max},k,d}, \Hat{\rho}_{d}, t_{d} \right], \vect{\xi})
\end{equation}
containing the reweighted signal-to-noise ratio and phase for each harmonic, the re-weighted total signal-to-noise ratio, the time-of-arrival, and the parameters of the template. The parameters within the square brackets are recorded for each detector, $d$, in the set of detectors, $\{d\}$, that observed the trigger. In this work the detector network is $\{ H, L\}$ representing the LIGO Hanford and Livingston detectors respectively. The parameters with a subscript $k$ are recorded for each harmonic used for the template that observed the trigger.

In~\cite{Davies:2020tsx} a ranking statistic is discussed based on the log-ratio of the signal event-rate density and the noise event-rate density. While this statistic is not formally an ``optimal'' statistic, it is more sensitive than just using the reweighted signal-to-noise ratio and was used to search for compact binary mergers with PyCBC in the O3 observing run~\cite{Davies:2020tsx, LIGOScientific:2021djp, Nitz:2021zwj}. We wish to develop a version of that statistic, which can be applied to out precessing search. We do this using a slightly modified version of that detection statistic given by
\begin{equation}
    \label{eq:log_rate_ratio_harmonic}
    \begin{split}
    R(\vect{\kappa}) = & - \log A_{\{H, L\}} - \sum_d \log r_{n, d}(\Hat{\rho}, \vect{\xi}) \\
    & - \log p(\vect{\Omega}|N) + \log p(\vect{\Omega}|S)
    \end{split}
\end{equation}
where $A_{\{H, L\}}$ is the allowed time window for coincident triggers between the two detectors, $r_{n, d}(\Hat{\rho}, \vect{\xi})$ is the expected noise-rate density for a trigger with re-weighted signal-to-noise ratio $\Hat{\rho}$ and template parameters $\vect{\xi}$ in detector $d$, finally, $p(\vect{\Omega}|N)$ and $p(\vect{\Omega}|S)$ are the likelihoods of a trigger having a set of extrinsic parameters $\vect{\Omega}$ given that it is a noise trigger or signal respectively.

For this work we have not included the relative sensitivity for the detector network and template included in~\cite{Davies:2020tsx}. This term would need to be updated to take into account the sensitivity of the different harmonics and the relative amplitudes of each. This will be required if extending to a larger network of detectors so that the detection statistic is comparable across different combinations of detectors. However, as we are limiting ourselves to two detectors we do not consider this term.

The factor $A_{\{H, L\}}$ can be calculated in the same way as for the aligned-spin search~\cite{Davies:2020tsx}, while the other terms will require modification. Let's consider first the noise-rate density for a single detector, $r_{n, d}(\Hat{\rho}, \vect{\xi})$. $r_{n, d}(\Hat{\rho}, \vect{\xi})$ is estimated using a decaying exponential function. This model is parameterised by two parameters, the rate of triggers above the threshold $\mu_n$ and the slope of the exponential $\alpha$. These parameters are fit using a maximum likelihood method for each template and then averaged for templates with similar intrinsic parameters. Looking at figure~\ref{fig:harmonic_singles} we see that the distribution of single detector triggers will still be well approximated by a decaying exponential in the case of the harmonic search. When searching with the harmonic template bank the rate of noise triggers will increase with the number of harmonics used. When averaging the values of $\mu_n$ and $\alpha$ we must therefore ensure that we do not average over templates that use different numbers of harmonics. By doing this the factor $r_{n, d}(\Hat{\rho}, \vect{\xi})$ will naturally be larger for templates with more harmonics in order to account for the increased rate of noise triggers. When extending from two to three harmonics we saw a decrease in the sensitivity across the full parameter space due to the increase in the noise-rate when including an additional harmonic. However, only a fraction of the template in the bank use all three harmonics. By averaging the parameters of $r_{n, d}(\Hat{\rho}, \vect{\xi})$ separately for different values of $N$ we should be able to increase the maximum number of harmonics in use without increasing the noise-rate for templates using less than the maximum number of harmonics.

The term $p(\vect{\Omega}|S)$ gives the likelihood of a signal having a set of extrinsic parameters $\vect{\Omega}$. In the case of the aligned-spin search the extrinsic parameters include the estimated amplitude, phase and time-of-arrival for each detector in the network. The likelihood is then estimated using a histogram of the expected amplitude ratios, phase differences and time differences with respect to a chosen reference detector. The histogram is generated by drawing sky positions and orientations isotropically and calculating the expected amplitudes ratios, phase differences and time differences across the network. These values are binned and the histogram is then smoothed using a Gaussian kernel in order to account for the uncertainty in the measured parameters.

When using the harmonic template bank the number of extrinsic parameters is increased as we now have an estimated amplitude and phase for each harmonic within each detector. The ideal solution would be to calculate $p(\vect{\Omega}|S)$ using the full set of these parameters. This would also have the effect of removing or down-weighting much of the unphysical parameter space introduced by maximising independently over $A_k$ and $\phi_k$. However, this introduces many technical difficulties. First, for a template with $N$ harmonics in $D$ detectors a histogram including the full set of extrinsic parameters would include $(2D + 1) (N-1)$ dimensions, for three harmonics in two detectors this already gives a seven dimensional parameter space. Increasing to a network of three detectors would produce a fourteen dimensional parameter space, making the use of these histograms computationally impractical for use within a search.

Secondly, in the case of aligned-spin templates the amplitude ratios and phase difference between detectors are only dependent on the extrinsic parameters of the binary, ignoring differences in power spectral density between detectors. This means that we can generate a single histogram that can be used for all templates within the bank. In the case of the harmonic template bank this is still the case when comparing a single harmonic across multiple detectors, however, the relative amplitudes between different harmonics will depend on the parameter $b$. We would therefore require a set of histograms covering different values of $b$ in order to properly account for the uncertainty due to noise. This is complicated further by the fact that $b$ will change as the binary evolves and the amplitude ratio will therefore depend on the value of $b$ averaged over the sensitive window of detector. As the frequency of each harmonic is higher than the last the average factor of $b$ measured between each subsequent harmonic may be different. In order to compare amplitudes across different harmonics it will be important to study the size of this effect.

As a first step we will therefore compare the amplitudes and phases for a single harmonic across the two detectors in our network, in this case the harmonics in the two detectors will have the same value of $b^k$ and the amplitude ratio will not depend on the value of $b$, removing the dependence on the intrinsic parameters of the template. We choose the harmonic with the highest signal-to-noise ratio to be used as the reference and compare the amplitude ratio, phase difference and time difference with respect to the same harmonic in the second detector. This will require three histograms to be generated when using a maximum of three harmonics. The required histograms can be generated using the same method as the aligned-spin case using equations \ref{eq:harmonic_dk} and \ref{eq:harmonic_phik} to calculate the expected amplitude ratios and phase differences. In this case the likelihood $p(\vect{\Omega}|N)$ for noise triggers will be assumed to be uniform as in the aligned-spin search.

\begin{figure}
    \centering
    \includegraphics[width=0.45\textwidth]{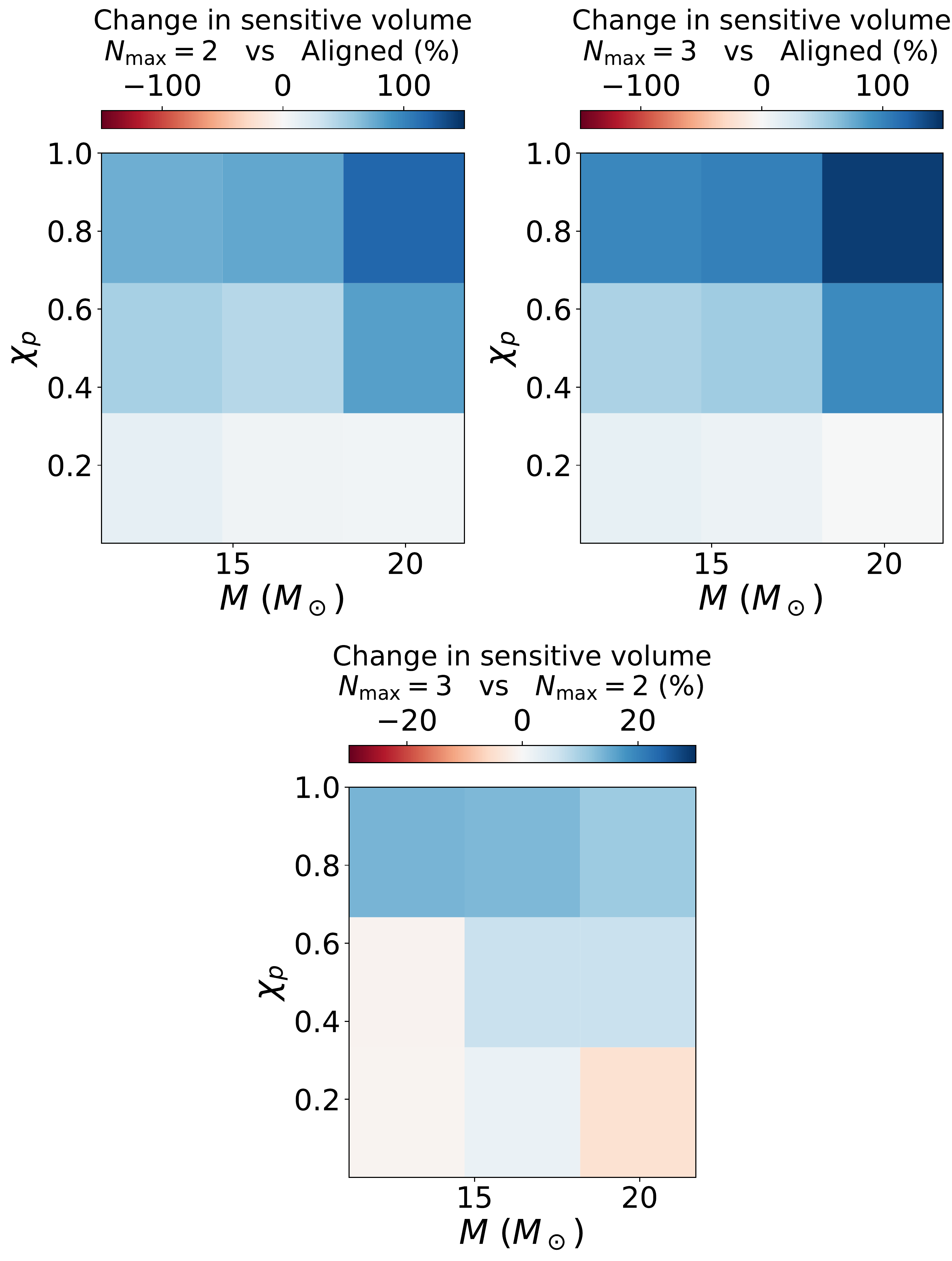}
    \caption{The change in sensitive volume of the search due to the harmonic template bank when using the log signal-rate noise-rate ratio statistic. The top left and top right panels show the change in sensitive volume from the aligned-spin template bank to the harmonic template bank with a maximum of two and three harmonics respectively. The bottom panel shows the change in the sensitive volume between the harmonic template bank with a maximum of two and three harmonics respectively.}
    \label{fig:rate_vt}
\end{figure}

We can then calculate the coincident ranking statistic using equation \ref{eq:log_rate_ratio_harmonic} and produce updated false-alarm rates. We apply this coincident ranking statistic using the same template bank and simulated signals as the previous sections. For comparison we use the full ranking statistic described in \cite{Davies:2020tsx} when performing the search using the aligned-spin template bank.

\subsection{Results with improved statistic}

Figure \ref{fig:rate_vt} shows the change in sensitivity when applying this new ranking statistic. In the case of the two harmonic template bank we see a similar improvement over the aligned-spin template bank as we did before introducing the new ranking statistic.

However, when using the harmonic template bank with a maximum of three harmonics we now see an improved sensitivity over using a maximum of two harmonics and do not see a sensitivity loss for signals with small values of $\chi_p$. For systems with $\chi_p > 0.67$ and total mass larger than $17.5 M_{\odot}$ we see an increase in sensitivity of $\sim 100\%$. This demonstrates that with the appropriate choice of ranking statistic we can account for the increased noise rate caused by introducing additional harmonics while gaining sensitivity through the improved match with signals in the data.

It will certainly be possible to improve upon this ranking statistic by using a more complete treatment of $p(\vect{\Omega}|S)$ in the future. We have restricted ourselves to comparing a single harmonic across each detector. It would be favourable to extend this to a larger set of harmonics in order to remove some of the unphysical freedom introduced by the maximisation of $A_k$ and $\phi_k$. One possibility is to first compute $p(\vect{\Omega}|S)$ in each detector separately and include it as a signal-consistency test in order to reduce the rate of single detector triggers, before subsequently testing a reduced set of harmonics across the detector network. This would require a maximum of four phase differences and four amplitude ratios for the single detector case if using all five harmonics, making the dimensionality more manageable. However, this will still require careful treatment of the averaged value of $b$ between different harmonics.

No significant signals were observed in the stretch of data analysed during this test, with the most significant foreground trigger producing a false-alarm rate of $\sim 200\,\text{yr}^{-1}$, which is consistent with the values we expect from noise triggers for this amount of data.

\section{Conclusion and outlook}

In this work we have demonstrated a method for searching for precessing binaries using a template bank utilising a harmonic decomposition of precessing signals. We have shown that given the natural hierarchy of the harmonics we can often use less than the full set of five harmonics and have demonstrated a method for selecting the appropriate number of harmonics to be used for each template within a bank.

By introducing extra parameters to our template model we do not only increase the observed signal-to-noise ratio for signals but increase the rate of noise triggers as well. We have shown that by using an appropriate ranking statistic this can be mitigated and an effective search can be run using the first three harmonics, with a $\sim 100\%$ increase in the sensitive volume compared to the aligned-spin search when considering binaries with $\chi_p > 0.67$ and total mass larger than $17.5 M_{\odot}$.

Two main improvements can be made to the current methods presented in this paper. We could construct a larger template bank in order to achieve a more complete coverage of the targeted parameter space, improving the signal-to-noise ratio of the observed signals. We could also extend the estimation of $p(\vect{\Omega}|S)$ to use more than one harmonic, allowing us to reduce the unphysical freedom caused by the maximisation of $A_k$ and $\phi_k$.

However, the results we present here demonstrate for the first time how a search on Advanced LIGO, Virgo and KAGRA data can be performed using precessing compact binary mergers as waveform filter templates and will acheive a significant sensitivity improvement for such signals. With scope for improving this method further we believe that this method can be the key to uncovering a potential population of binaries with strong precessional effects. 

\section{Acknowledgements}

We thank Patricia Schmidt for insightful and helpful comments on this work while reviewing an earlier draft of the text, which appeared in CM's thesis.
We thank Stephen Fairhurst for motivating why the ``two harmonic" approximation would be right thing to use for precessing searches, even if it turns out that two harmonics may not be enough. We thank Mark Hannam for discussions concerning the inconsistencies with the \texttt{IMRPhenomXP} waveform model.
CM acknowledges the DISCNet Center for Doctoral Training for support.
CH thanks the UKRI Future Leaders Fellowship for support through the grant MR/T01881X/1. IH thanks the STFC for support through the grants ST/T000333/1 and ST/V005715/1. The authors are also grateful for computational resources provided by LIGO Laboratory and supported by National Science Foundation Grants PHY-0757058 and PHY-0823459.

This research made use of data, software and/or web tools obtained from the Gravitational Wave Open Science Center (\href{https://www.gw-openscience.org}{https://www.gw-openscience.org}), a service of LIGO Laboratory, the LIGO Scientific Collaboration and the Virgo Collaboration. LIGO is funded by the U.S. National Science Foundation. Virgo is funded by the French Centre National de Recherche Scientifique (CNRS), the Italian Istituto Nazionale della Fisica Nucleare (INFN) and the Dutch Nikhef, with contributions by Polish and Hungarian institutes. This material is based upon work supported by NSF's LIGO Laboratory which is a major facility fully funded by the National Science Foundation.

\appendix
\section{Highly precessing injection set}
\label{sec:app_highboth_injections}

\begin{figure}
    \centering
    \includegraphics[width=0.45\textwidth]{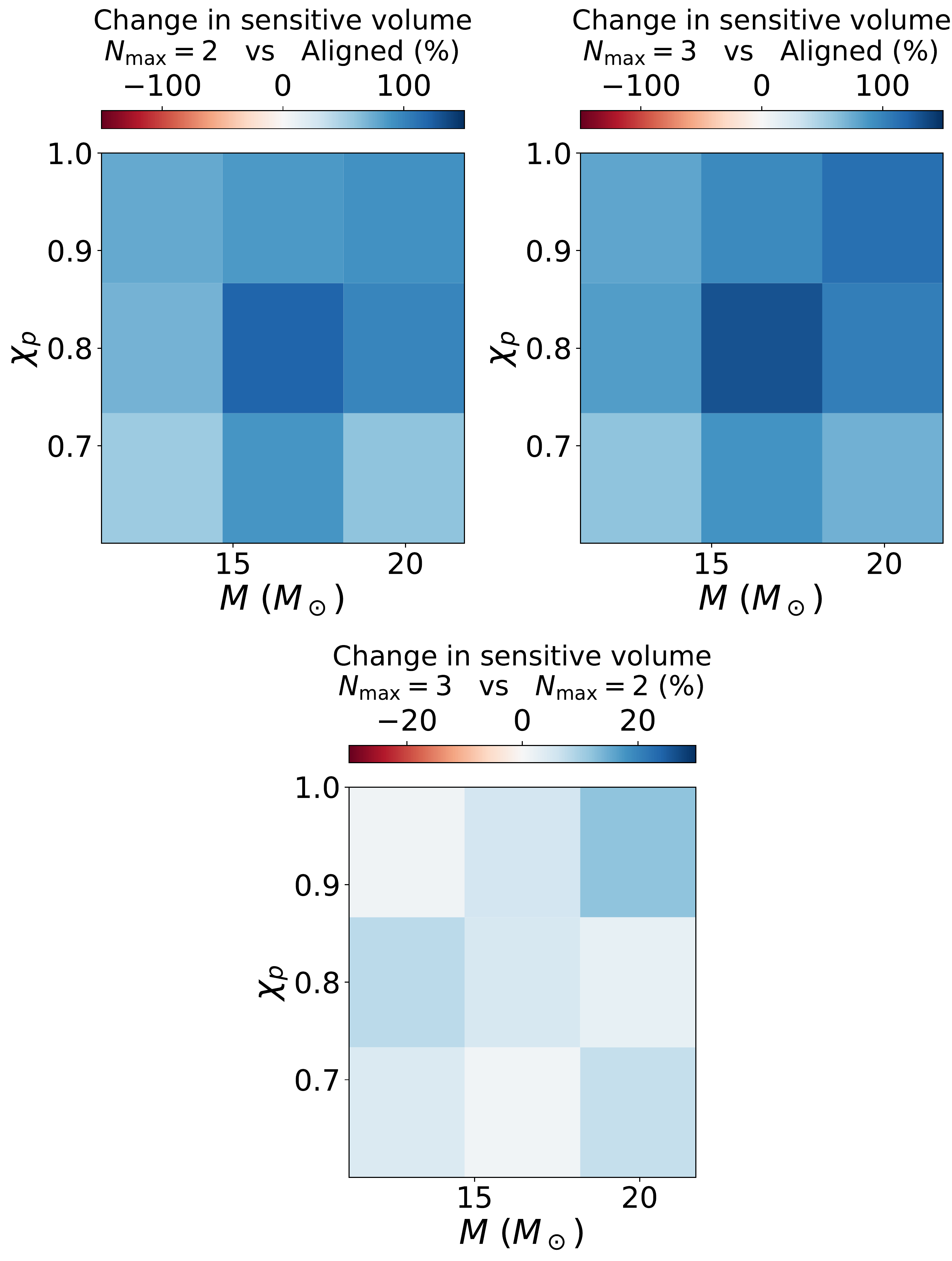}
    \caption{Same as Fig.~\ref{fig:rate_vt} but using a set of highly precessing simulated signals to estimate the sensitive volume.}
    \label{fig:rate_vt_highboth}
\end{figure}

In Sec.~\ref{sec:harmonic_coinc}, we added a set of simulated signals into real gravitational-wave strain data to evaluate the sensitivity of our search. It was demonstrated that the sensitive volume improved by $\sim 50\%$, and $\sim 60\%$ on average when using a precessing template bank limited to 2 and 3 harmonics respectively.

We also show the improvement in sensitive volume for a set of simulated signals that are highly precessing: spins almost entirely within the plane of the binary. In comparison to the previous set, the new set of simulated signals modified only the primary spin distribution: the spin magnitudes were drawn uniformly between 0.7 and 0.99 and spin orientations drawn isotropically between $|\cos ( \vect{\Hat{S}_{1}} \cdot \vect{\Hat{L}} )| < 0.3$.

We see that, when using an appropriate ranking statistic, the sensitive volume improves by $\sim 80\%$, and $\sim 90\%$ on average when using a precessing template bank limited to 2 and 3 harmonics respectively, compared to an aligned-spin search. We also see that the 3 harmonic search out performs the 2 harmonic search for all $\chi_{p}$ and $M$.

\section{Waveform inconsistencies reducing the bank size}
\label{sec:app_phenomx_inconsistencies}

\begin{figure*}
    \centering
    \includegraphics[width=0.98\textwidth]{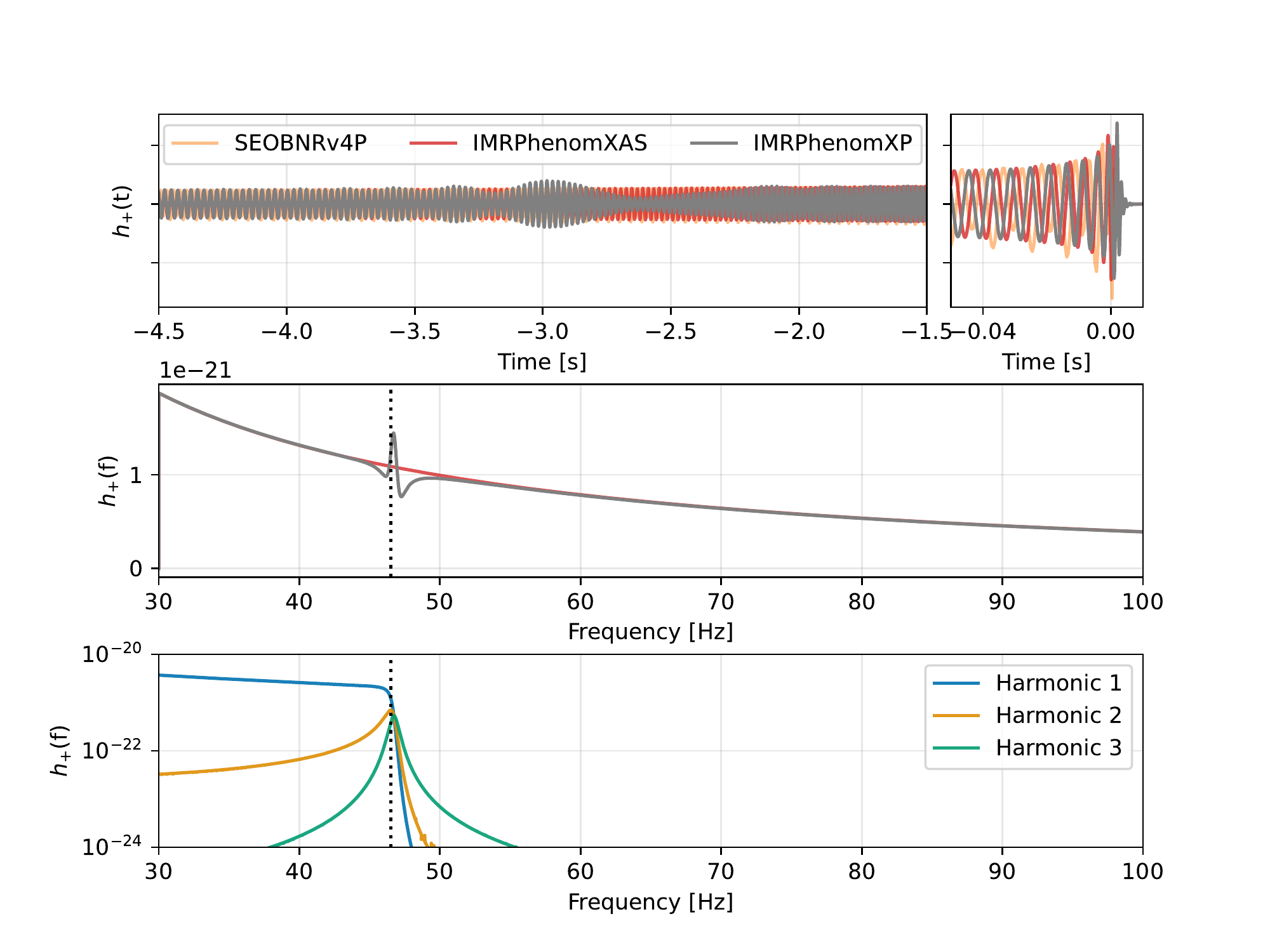}
    \caption{Comparison of the gravitational-waves produced using different waveform models for a binary with masses $13.4\, M_{\odot}$ and $1.3\, M_{\odot}$, and component spins $\vect{s}_1 = (0.0003, -0.0002, -0.5)$, $\vect{s}_2 = (-0.02, -0.1, -0.2)$, viewed at an inclination angle of 1.5 radians. The top two panels show the gravitational-wave in the time-domain (\emph{Left}: focusing on inspiral, \emph{Right}: focusing on the merger and ringdown), the middle panel shows the gravitational-wave in the frequency-domain, and the bottom panel shows the harmonic decomposition of the gravitational-wave. The vertical black dotted line in the middle and bottom panels corresponds to 47 Hz.}
    \label{fig:waveform_inconsistency}
\end{figure*}

After constructing the template bank through stochastic bank generation techniques (see Sec.~\ref{sec:harmonic_bank} for details), we removed 1\% of filter waveforms due to inconsistencies with the \texttt{IMRPhenomXP} waveform model. In our testing, we found that a several filter waveforms had non-physical artifacts, particularly at low frequencies. This caused issues when deconstructing the waveform into the harmonic decomposition (see Sec.~\ref{sec:harmonics} for details), and consequently, caused a background of large SNR events which biased our results.

We found that most of the inconsistent waveforms had low in-plane spins. In Fig.~\ref{fig:waveform_inconsistency} we specifically show one of filter waveforms that caused issues; this waveform was generated for a binary with masses $13.4\, M_{\odot}$ and $1.3\, M_{\odot}$, with component spins $\vect{s}_1 = (0.0003, -0.0002, -0.5)$, $\vect{s}_2 = (-0.02, -0.1, -0.2)$, viewed at an inclination angle of 4.8 radians. For comparison we also show the gravitational-waves produced for the same binary configuration but with alternative models: \texttt{SEOBNRv4P}~\cite{Ossokine:2020kjp}, which constructs precessing signals through the effective one-body approach~\cite{Buonanno:1998gg}, and the \texttt{IMRPhenomXAS} model, which constructs aligned-spin signals through the phenomenological approach. When generating the gravitational-wave with the \texttt{IMRPhenomXAS} model, we consider the aligned-spin projection of the binary (all in-plane spins reduced to exactly 0). From the frequency-domain, we can clearly see that the gravitational wave produced from the \texttt{IMRPhenomXP} waveform model has a non-physical artifact at 47Hz. Since this specific binary has $\chi_{p} = 0.008 \ll 1$, it should closely resemble an aligned-spin signal. As expected, wee see excellent agreement between the \texttt{IMRPhenomXAS} and \texttt{SEOBNRv4P} waveform models, suggesting that the issue is specific to the \texttt{IMRPhenomXP} model.

When deconstructing this specific binary into the harmonic decomposition, the non-physical artifact at 47Hz caused the leading harmonic to drop in amplitude from $O(10^{-21})$ to $O(10^{-29})$, and the $5^{\mathrm{th}}$ harmonic to increase in amplitude from $(10^{-27})$ to $O(10^{-21})$ to compensate. This means that when using only the leading two, or leading three harmonics, in the search analysis presented in Section~\ref{sec:harmonic_coinc}, the reconstructed precessing waveform terminated at 47Hz.

In order to construct a bank that neglected these inconsistent filter waveforms, we removed the waveforms that had either: (a) $\chi_{p} < 0.05$ (meaning that they closely resemble aligned-spin binaries) and a match between the \texttt{IMRPhenomXP} and \texttt{IMRPhenomXAS} $< 0.85$, or (b) a leading harmonic that was initially larger than the $5^{\mathrm{th}}$ harmonic, and then switching to having the $5^{\mathrm{th}}$ larger than the leading harmonic. These constraints removed $\sim 3700$ filter waveforms from a total of $360,000$.

\bibliographystyle{utphys}
\bibliography{main.bib}
\end{document}